\documentclass[aps,pra,reprint,groupaddress,superscriptaddress, showpacs]{revtex4-1}

\usepackage{mathtools}
\usepackage{amsfonts}
\usepackage{latexsym}
\usepackage{relsize}
\usepackage{mathrsfs}

\DeclareMathAlphabet{\mathbbold}{U}{bbold}{m}{n}

\usepackage{euscript}%changes caligraphic font, see http://
\usepackage{amssymb}
\usepackage{graphicx}
\usepackage{amsmath}
\usepackage{amsbsy}

\usepackage{amsthm}

\usepackage{bbm}
\usepackage{bm}
\usepackage{epsfig}
\usepackage{epstopdf}
\usepackage{dsfont}

\usepackage[colorlinks]{hyperref}
\usepackage[figure,table]{hypcap}

\usepackage{enumerate}

\hypersetup{
	bookmarksnumbered,
	pdfstartview={FitH},
	citecolor={darkgreen},
	linkcolor={darkred},
       linktoc={page},
	urlcolor={darkblue},
	pdfpagemode={UseOutlines}}

\usepackage{color}

\definecolor{darkgreen}{RGB}{40,130,40}
\definecolor{darkblue}{RGB}{0,0,190}
\definecolor{darkred}{RGB}{238,0,0}
\usepackage{soul}

\usepackage{float}
\usepackage{algorithm}

\floatname{algorithm}{Protocol}

\def\EQ#1{\begin{eqnarray}#1\end{eqnarray}}
\newcommand{\djj}{d\kern-0.4em\char"16\kern-0.1em}

\newcounter{lem}

\newtheorem{prop}{Proposition}\def\PRO{\begin{prop}}\def\ORP{\end{prop}}
\newtheorem{coro}{Corollary}\def\COR{\begin{coro}}\def\ROC{\end{coro}}
\newtheorem{theo}{Theorem}\def\TH{\begin{theo}}\def\HT{\end{theo}}
\def\TH{\begin{theo}}\def\HT{\end{theo}}
\newtheorem{defi}[prop]{Definition}\def\DE{\begin{defi}}\def\ED{\end{defi}}
\newtheorem{lemme}[lem]{Lemma}\def\LE{\begin{lemme}}\def\EL{\end{lemme}}

\def\S{\mathcal{S}}

\def\ket#1{\left| #1 \right\rangle}
\def\bra#1{\left\langle #1 \right|}
\def\dm#1{\left|#1 \right\rangle \left\langle #1 \right|}

\def\ve#1{\mathbf{#1}}

\usepackage{amsmath,amsfonts,amssymb}
\usepackage{wrapfig}
\usepackage{graphicx}
\usepackage{bbm}
\usepackage{graphics}

\def \beq {\begin{equation}}
\def \eeq {\end{equation}}
\def \ba {\begin{eqnarray}}
\def \ea {\end{eqnarray}}

\begin{document}
\title{Quantum-enhanced machine learning}
\date{\today}
\author{Vedran Dunjko}
\email{vedran.dunjko@uibk.ac.at}
\address{Institut f\"{u}r Theoretische Physik, Universit{\"{a}}t Innsbruck, Technikerstra{\ss}e 21a, A-6020 Innsbruck, Austria}

\author{Jacob M. Taylor}
\email{jmtaylor@umd.edu}
\address{Joint Quantum Institute, National Institute of Standards and Technology, Gaithersburg, MD 20899 USA} 
\address{Joint Center for Quantum Information and Computer Science, University of Maryland, College Park, MD 20742 USA}
\author{Hans J. Briegel}
\email{hans.briegel@uibk.ac.at}
\address{Institut f\"{u}r Theoretische Physik, Universit{\"{a}}t Innsbruck, Technikerstra{\ss}e 21a, A-6020 Innsbruck, Austria}
\begin{abstract}
The emerging field of quantum machine learning has the potential to substantially aid in the problems and scope of artificial intelligence. This is only enhanced by recent successes in the field of classical machine learning. In this work we propose an approach for the systematic treatment of machine learning, from the perspective of quantum information. Our approach is general and covers all three main branches of machine learning: supervised, unsupervised and reinforcement learning. While quantum improvements in supervised and unsupervised learning have been reported, reinforcement learning has received much less attention. Within our approach, we tackle the problem of quantum enhancements in reinforcement learning as well, and propose a systematic scheme for providing improvements. As an example, we show that quadratic improvements in learning efficiency, and exponential improvements in performance over limited time periods, can be obtained for a broad class of learning problems.  
\end{abstract}
\maketitle

%WordCountBegin; orig=3155; 3354; 199
\textit{Introduction.--} 
The field of artificial intelligence (AI) has lately had remarkable successes, especially in the area of machine learning \cite{2015_Nature, 2015_Science}. A recent milestone, until recently believed to be decades away -- a computer beating an expert human player in the game of Go \cite{2016_Silver} -- clearly illustrates the potential of learning machines. In parallel, we are witnessing the emergence of a new field: quantum machine learning (QML), which has a further, profound potential to revolutionize the field of AI, much like quantum information processing (QIP) has influenced its classical counterpart \cite{2000_NC}. 

The evidence for this is already substantiated with improvements reported in classification and clustering \cite{2014_Wittek,2013_Lloyd, 2013_Aimeur, 2014_Rebentrost} problems. 
Such tasks are representative of two of the three main branches of {machine learning}.  The first, supervised learning, considers the problem of learning the conditional distribution $P(y|x)$ (e.g., a function $y=f(x)$) which assigns labels $y$ to data $x$ (i.e. classifies data), {based on correctly-labeled examples, called \emph{the training set},}  provided from a distribution $P(x,y).$ 
The second, unsupervised learning, uses samples to identify a structure in a distribution $P(x),$ e.g., identifies clusters. 
{The quantum analog of the first task corresponds to a tomography-type problem where conditional states $\rho^x_Y$(states of a partition of a system, given a measurement outcome of another partition) }
{should be reconstructed from the measurement statistics of the joint state $\rho_{XY},$ which encodes the  distribution $P(x,y).$ The unsupervised case is similar.}

The third branch, reinforcement learning (RL) constitutes an interactive mode of learning, and is more general. {Here the \emph{learning agent} (or learning algorithm) learns how to behave correctly through the use of reinforcement signals -- rewards, or punishments.} RL has been less investigated from {a quantum information} perspective, although some results have been reported \cite{2005_Dong,2014_Paparo}.

The key question of how quantum processing can help in learning requires us to clarify what constitutes a \emph{good} learning model. This can be involved, but two characteristics are typically considered. The first is the \emph{computational complexity} of the algorithm of the learner. The second, \emph{sample complexity}, is standard for supervised learning, and quantifies how large the training set has to be, for the algorithm to learn the distribution $P(y|x)$. {That is, in a tomography context, it counts the number of copies of $\rho_{XY}$ required until the learning algorithm  can reconstruct the states $\rho^x_Y$  to desired confidence.}

In RL, sample complexity is usually substituted by \emph{learning efficiency} -- the number of interaction steps needed for the agent to learn to obtain the rewards with high probability.
The recent results in QML have focused on improving computational complexity \cite{2014_Wittek,2013_Lloyd, 2013_Aimeur, 2014_Rebentrost,2014_Paparo}, with only few recent works considering sample complexity aspects \cite{2016_Wolf}  or  {supervised computational learning} \cite{2001_Servedio,2006_Atici}. 
However, the broader question of how, and to what extent, AI can ultimately benefit from quantum mechanics, in general learning settings, remains largely open.

In this work we address this question, with emphasis on the more general, and less explored, RL setting. 
We propose a paradigm for considering QML, which allows us to better understand its limits and its power.
Using this, we present a schema for identifying settings where quantum effects can help. To illustrate how the schema works, we provide a method for achieving quantum improvements (polynomial in the required number of interaction rounds and exponential improvements in success rate) in many RL settings. 

\textit{A paradigm for QML.--} 
All three {learning} settings fit in the paradigm of so-called learning agents \cite{2003_Russel}, {standard in  the field of artificial intelligence}.
Here we consider a learning agent $A$ ({equivalently a learning program $A$}) which interacts with an unknown environment $E$  ({the so-called \emph{task environment}, or \emph{problem setting}}) via the exchange of messages, interchangeably issued by $A$ (called \emph{actions} $\mathcal{A} = \{ a_i\}$) and $E$ (called \emph{percepts} $\S=\{s_j\}$).
In the quantum extension, these sets become Hilbert spaces, $\mathcal{H}_\mathcal{A} = \textup{span} \{\ket{a_i}\},$ $\mathcal{H}_\mathcal{S} = \textup{span} \{\ket{s_i}\},$ and form orthonormal bases. The percept and action states, and their mixtures, are referred to as classical states. 
 Any figure of merit $Rate(\cdot)$ of the performance of an agent $A$ in $E$ is a function of the \emph{history of interaction} $\mathbf{H} \owns h = (a_1, s_1,\ldots),$ collecting the exchanged percepts and actions. 
The history of interaction is thus the central concept in learning. The correct quantum generalization of the history is not trivial, and we will deal with this momentarily.

 If either $A$ or $E$ are stochastic, the interaction of $A$ and $E$ is described by a 
 distribution over histories (of length $t$), denoted by $A \leftrightarrow_t E$.
  Most figures of merit are then extended to such distributions by convex-linearity.
 
To recover, e.g., supervised learning in this paradigm, {take $E$ to be characterized by the distribution $P(x,y),$ where the agent is given the training set -- $n$ labeled data points (pairs $(x,y)$) sampled from $P(x,y)$  -- as the first $n$ percepts}. After this, the agent is to {respond with the correct labels as actions (responses) to the presented percepts, which are now the \emph{unlabeled} data-points $x$. }
{Reinforcement learning is naturally phrased as such an agent-environment interaction,  where the percept space also contains the reward.} We denote the percept space including the reward status as $\bar{\S}$ (e.g., if rewards are binary then $\bar{\S} = \S \times \{0,1\}$). 

Formally, the agent-environment paradigm is a two-party interactive setting, and thus convenient for a quantum information treatment of QML. All the existing results group into four categories \cite{2006_Brassard}: $CC,CQ,QC$ and $QQ$, depending on whether the agent (first symbol) or the environment (second symbol) are classical ($C$)  or quantum ($Q$).
This classification is reminiscent to, but should not be confused with, the classification of quantum computational universality \cite{2007_Nest} where $C/Q$ specify whether the input/outputs of a quantum computation are classical.
The $CC$ scenario covers classical machine learning. The $CQ$ setting asks how classical {learning} techniques may aid in quantum tasks, such as quantum control \cite{2014_Zahedinejad, 2010_Wiseman}, quantum metrology \cite{2013_Lovett}, adaptive quantum computing \cite{2015_Tiersch} and the design of quantum experiments \cite{2016_Kren}. 
Here we deal with, for example,  nonconvex or nonlinear optimization problems arising in quantum experiments, tackled by {machine learning} techniques.
$QC$ corresponds to quantum variants of learning algorithms \cite{2014_Paparo, 2014_Schuld, 2013_Aimeur} facing a classical environment. 
Figuratively speaking, this studies the potential of a learning robot, enhanced with a ``quantum chip''. 
 In $QQ$ settings, the focus of this work, both $A$ and $E$ are quantum systems. Here, the interaction can be fully quantum, and even the question of what it means ``to learn'' becomes problematic as, for instance, the agent and environment may become entangled. 

\begin{figure}[!h]
\includegraphics[width=0.47\textwidth, trim=0cm 12.9cm 0cm 12.9cm,clip=true]{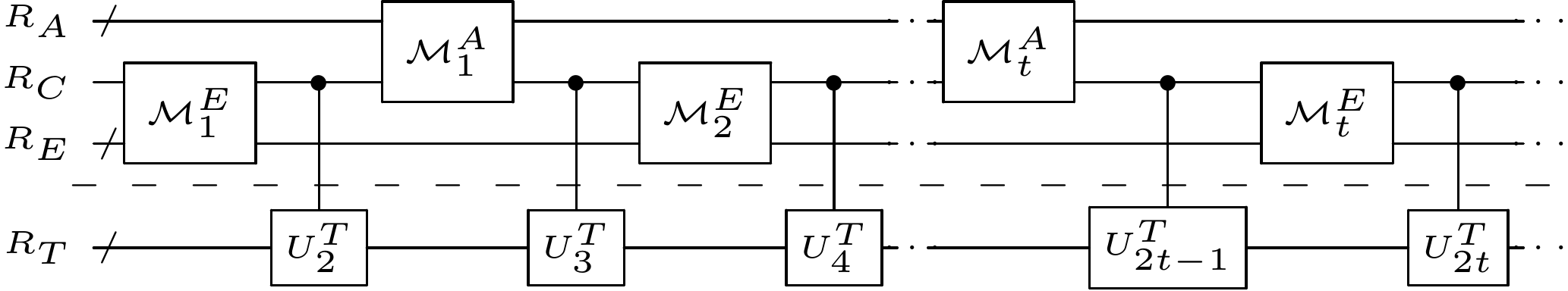}
\caption{\label{fig2}
Tested agent-environment interaction. In general, each map of the tester $U^T_k$ acts on a fresh subsystem of the register $R_T$, which is not under the control of the agent, nor of the environment. The crossed wires represent multiple systems.}
\end{figure}

\textit{Framework.--} Since learning constitutes a two-player interaction, standard quantum extensions can be applied: the action and percept sets are represented by the aforementioned Hilbert spaces $\mathcal{H}_\mathcal{A}, \mathcal{H}_\mathcal{S}$. 
The agent and the environment act on a common communication register $R_C$ (capable of representing both percepts and actions). Thus, the agent (environment) is described as a sequence of completely positive trace-preserving (CPTP) maps $\{\mathcal{M}_A^t\}\ (\{\mathcal{M}_E^t\})$-- one for each time-step -- which acts on the register $R_C$, but also a private register $R_A$ ($R_E$) which constitutes the internal memory of the agent (environment). This is illustrated in Fig. \ref{fig2} above the dashed line.

The central object characterizing an interaction, namely its history, is, for the quantum case, generated by performing periodic measurements on $R_C$ in the classical (often called computational) basis. The generalization of this process for the quantum case is a \emph{tested interaction}: we define the \emph{tester} as a sequence of controlled maps of the form
 \EQ{
U^T_t \left( \ket{x}_{R_C} \otimes \ket{\psi}_{R_T} \right) = \ket{x}_{R_C} \otimes U^{x}_t\ket{\psi}_{R_T} \label{tester}\nonumber
} 
where $x \in \mathcal{S} \cup \mathcal{A},$ and $\{ U^{x}_t\}_x$ are unitary maps acting on the tester register $R_T$, for all steps $t$. The history, relative to a given tester, is defined to be the state of the register $R_T$. A tested interaction is shown in Fig. \ref{fig2}.

The restriction that testers are controlled maps relative to the classical basis guarantees that, for any choice of the local maps $U^{x}_T$, the interaction between classical $A$ and $E$ remains unchanged. A \emph{classical tester} copies the content of $R_C$ relative to the classical basis, which has essentially the same effect as measuring $R_C$ and copying the outcome. In other words, the interface between $A$ and $E$ is then classical.
It can be shown that, in the latter case, for any quantum agent and/or environment there exist classical $A$ and $E$ which generate the same history under any tester (see the Appendix for details). In other words, classical agents can, in $QC$ settings and, equivalently, in classically tested $QQ$ settings, achieve the same performance as quantum agents, in terms of any history-dependent figure of merit. Thus, the only improvements can then be in terms of computational complexity.  

\textit{Scope and limits of quantum improvements.--} 
What is the ultimate potential of quantum improvements in learning?  In the $QC$ and classically tested settings, we are bound to computational complexity improvements, which have been achieved in certain cases.
Improvements in learning efficiency require special type of access to the environments, which is not fully tested.
Exactly this is done in \cite{2013_Lloyd, 2014_Rebentrost}, for the purpose of improving computational complexity, with great success, as the improvement can be exponential. There, the classical source of samples is substituted by a quantum RAM \cite{2008_Giovannetti} architecture, which allows for the accessing of many samples in superposition.
Such a substitution comes naturally in (un)supervised settings, as the basic interaction comprises only two steps and is memoryless -- the agent requests $M$ samples, and the environment provides them.  However, in more general settings, environments are ill-suited for such quantum parallel approaches: in general, the environment stores all the actions of the agent it in its memory, never to return them again. 
This effectively breaks the entanglement in the agent's register $R_A$, and prohibits all interference effects.
 Nonetheless, for many environmental settings, it is still possible to `dissect' the maps of the environment, and to provide oracular variants, which we can use to help the agent learn.

\textit{An approach to quantum improvements in reinforcement learning.--} 
This brings us to our schema for improving RL agents. First, given a classical environment $E,$ we define \emph{fair unitary oracular equivalents} $E^q$. Here, fair is meant in the same sense as quantum oracles of boolean functions are fair analogues of classical boolean functions --
$E^q$ should not provide more information than $E$ under classical access, which is guaranteed, e.g., when $E^q$ is realizable from a \emph{reversible} version of $E$. 
Second, as access to any \emph{quantum environment} $E^q$ cannot generically speed-up all aspects of an interaction (e.g., while quantum walks can find target vertices faster, the price is that the traversed path is undefined), we identify particular environmental properties which can be more efficiently ascertained using $E^q$, and which are relevant for learning. 
Third, we construct an improved agent which uses the properties from the previous points.
We now illustrate our approach on a restricted scenario, for the ease of presentation, and show how the examples generalize later.

\textit{Application of the framework.--} 
Given any task environment, {we can separately consider the map which specifies the next percept the environment will present-- in general, a stochastic function $f_E: \mathbf{H} \rightarrow \mathcal{S},$ mapping elapsed histories onto the next percept -- and the reward function. function. The latter is described as the map $\Lambda: \mathbf{H} \times \S \rightarrow \bar{\S}$ which also depends on the history, and complements the percept by setting its reward status. }
In environments which are simple and strictly epochal (meaning the environment is re-set after $M$ steps and at most one reward is given), although the interaction is turn-based, it can be represented as sequences of $M-$step maps:
 \EQ{
\ket{a_1, \ldots, a_M} \longrightarrow \ket{s_1, \ldots, \overline{s_M}},\label{E-classical}
}
where the ``bar" on $s_M$ highlights that it includes a reward status. 
Moreover, in deterministic environments, the maps $f_E$ and $\Lambda$  only depend on the actions of the agent, as the percept responses are fixed. 
For such deterministic, simple strictly epochal environments, the construction of an appropriate oracle is dramatically simplified. The actions can be returned to the agent after each block of $M$ steps, as the next block is independent. { Moreover, using phase kick-back, the reward map can be modified (see the Appendix for details) such as to influence just the global phase of returned action states. This leads to the  ``phase-flip'' oracle realizing}
\EQ{
\ket{a_1, \ldots, a_M} \stackrel{E^q_{\textup{\textit{oracle}}}}{\longrightarrow} (-1)^{\Lambda(a_1, \ldots, a_M)} \ket{a_1, \ldots, a_M} \label{E-quantum},
}
{One use of this environment-specific oracle requires $M$ interaction steps.}
This constitutes the first step of our proposed schema.
Next, we focus on step two: obtaining a useful property of the environment, and identifying settings where it provably helps.
The constructed oracle points towards the use of Grover-type search to find rewarding action sequences. This alone suffices for improvements only in special environments where learning reduces to searching. 
We can do better by combining fast search with a classical learning model.
In canonical RL settings, what the agent learns (should learn!) is not a correct sequence of moves per se, but rather the correct association of actions given percepts. To illustrate this, imagine navigating a maze where the percepts encode correct directions of movement. If the correct association is learned, then the agent will perform well, even when the maze changes. Nonetheless, for the agent to learn the correct association, it first must encounter an instance of rewarding sequences, and here quantum access helps. Thus we aim at assisting in the exploration phase of the balancing act between exploration (trying out behaviors to find optima) and exploitation (reaping rewards by using learned information) characteristic for RL \cite{SuttonBarto98}. 
{This idea can be made fully precise by considering the class of environments where more successful exploration phases are guaranteed to lead to a better overall learning performance. Whether this is the case, however, also depends also on the learning model of the agent. Thus we identify agent-environment pairs, where such better performance in the past (in exploration) implies better performance in the future (on average), which we call \emph{luck-favoring settings}.}

More formally, consider environments $E,$ and agents $A,$ such that 
if $h_t$ and $h'_t$ are $t-$length histories, then $Rate(h_t)>Rate(h'_t)$ {(i.e. $h_t$ is a history with a better performance than $h'_t$)} \textup{implies} 
\EQ{
Rate(E(h_t)\! \leftrightarrow_T\! A(h_t)) \geq Rate(E(h'_t)\! \leftrightarrow_T\! A(h'_t)) \label{eq-LF},
} for some future period $T$. {Here $E(h)$ and $A(h)$ denote the environment and agent, respectively, which have undergone the history $h,$ (note that $A(h)$ and $A$ are, technically, different agents).}

We will say $A(h_t)$ is luckier than $A(h_t')$. {Such environment-agent pairs $(A,E)$, satisfying the formal conditions above, are thus luck-favoring,} and we may additionally specify the periods $t$ and $T$ for which the implication (\ref{eq-LF}) holds. This brings us to step three of the schema, given as a Theorem. 

\vspace{0.3cm}
\noindent\textbf{Theorem 1} \textit{
Let $E$ be a deterministic, strictly epochal environment. Then there exists an oracular variant $E^q$ of $E$, such that for any classical learning model $A$ which is luck-favoring relative to $E$, and figure of merit \emph{Rate} which is monotonically increasing in the number of rewards in the history, we can construct a quantum agent $A^q$ such that $A^q$, by interacting with $E^q$,  outperforms $A$ in terms of the figure of merit \emph{Rate} relative to a chosen tester.
}

\vspace{0.3cm}
This Theorem states that, in the restricted settings of determinisic epochal environments, it is possible to generically improve the learning efficiency of all learning agents, provided the environments are luck-favoring for those agents. We note that most reasonable learning models are luck-favoring relative to most typically considered task environments (see the Appendix for a longer discussion). 
In the statement of Theorem 1, we have omitted additional specifications pertaining to $t$ and  $T,$ but it should be understood that if the luck favoring property holds for $t$ and $T,$ then the improved performance holds relative to these periods.

\begin{figure}
\fbox{
\includegraphics[width=0.45\textwidth, trim=0cm 0.2cm 0cm 0.15cm,clip=true]{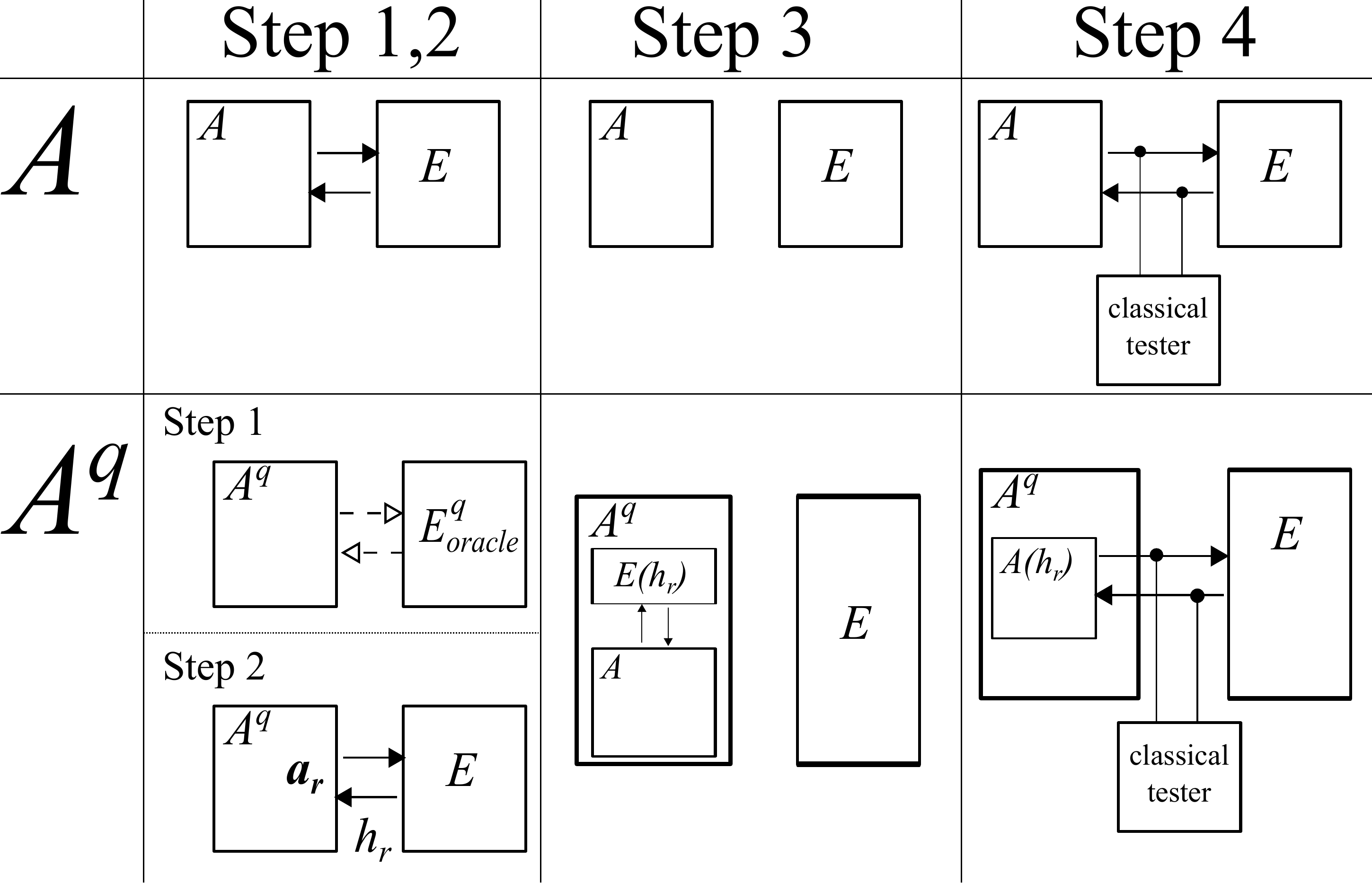}
}
\caption{\label{construction-main}
Differences between the interaction for $A$ and $A^q$. In Steps 1 and 2, $A^q$ uses access to $E^q_{oracle}$, for $O(t)$ steps, and obtains a rewarding sequence $h_r$. Step 3: $A^q$ simulates the agent $A$, and `trains' the simulation to produce the rewarding sequence. In Step 4, $A^q$ uses $A(h_{r})$ for the remainder of the now  classically tested interaction, with the classical environment $E$. 
}
\end{figure}

To prove Theorem 1 we construct $A^q$, given $A$. 
The construction is illustrated step by step in Fig. \ref{construction-main}, where for illustrative purposes, the classical interaction of agent $A$ is contrasted against the quantum interaction of agent $A^q$.
Step 1: $A^q$ will use the quantum oracle variant of $E$ ( $E^q_{oracle}$) for time $t \in O(\sqrt{|A|^M }),$ where $M$ is the epoch length, and $|A|$ is the number of the actions, to find a rewarding action sequence $\mathbf{a}_r$, using Grover search. During this period the interaction is untested, and the interaction is fully classically tested thereafter.
Step 2:  $A^q$ will play out one epoch by outputting actions from $\mathbf{a}_r$ sequentially, now with the classical environment, to obtain the responses of the environment (recall, $E^q_{oracle}$ cannot provide these), obtaining the entire rewarding history $h_r$. Thus far, $A^q$ used $O(M\sqrt{|A|^M })$ interaction steps. Step 3:  $A^q$ ``trains'' an internal simulation of $A$, simulating the interaction between $A$ and $E,$ and restarting the simulation until the history $h_r$ occurs (we assume such an occurrence has a non-zero probability).
This may require many internally simulated interactions, but no interaction with the real environment.
In Step 4, the internal simulation of $A(h_r)$ corresponds to the luckiest agent possible, and $A^q$ relinquishes control to it.

Finally, we consider what happens with $A$ during the same time periods. Unless additional information about the environment is given, in $O(t)$ steps $A$ has only an exponentially small ($O(\exp(-M \ln(|A|)/2))$) probability of having seen the rewarding sequence. Thus, the quantum agent is luckier than the classical, and in luck-favoring settings, this implies that $A^q$ will continue to outperform $A$ after the $t$ steps. 
The statement of Theorem 1 is not quantitative, due to the generality of the definition of luck-favoring settings. We can, however, trade off generality for exactness. If an agent $A$ employs a variant of $\epsilon-$greedy \cite{SuttonBarto98} behavior -- that is, it outputs the rewarding sequence (exploits) with probability $\epsilon$ and explores with probability $1-\epsilon$, then the ratio of the performances of $A^q$ and $A$ will be exponential in $M$: the constant reward probability $\epsilon$ of $A^q$ versus the exponentially diminishing  $O(\exp(-M \ln(|A|)/2)))$ of $A$ at step $t$. This exponential gap holds for time-scales $T \in O(t).$ However, the improvement in terms of learning efficiency (number of interaction steps) is quadratic. 

Our results achieve solid improvements using simple techniques, at the cost of restricting the task environments.
However, our example can be further generalized in two directions.

First, as long as the re-set occurs at step $M$, multiple and multi-valued rewards can also be handled by defining oracles which reversibly count the rewards. Highly rewarding sequences can then be found through quantum optimization techniques~\cite{1996_Durr}, as worked out in the Appendix.

Second, under stronger assumptions on $E^q$, using more involved quantum subroutines, we can deal with stochastic environments. For instance, in the setting with one reward per epoch, the oracle 
\EQ{
\ket{\mathbf{a}}\ket{{0}} \stackrel{\mathcal{U}_E}{\longrightarrow} \ket{\mathbf{a}}( \cos \theta_\mathbf{a} \ket{0} + \sin \theta_\mathbf{a} \ket{1})
}
where $\sin^2 \theta_\mathbf{a}$ is the probability of a reward, given the action sequence $\mathbf{a}$,
 can be constructed from a reversible implementation of the environment where randomness is represented as a subsystem of an entangled state (see the Appendix for details).

From here, by using phase kick-back and phase estimation
 the agent can realize the mapping
\EQ{
\ket{\mathbf{a}}\ket{\mathbf{0}} \rightarrow  \ket{\mathbf{a}}\ket{\tilde{\mathbf{\theta}}_\mathbf{a}}, 
}
where $\tilde{\mathbf{\theta}}_\mathbf{a}$ is an $l-$bit precision estimate of the reward probability as specified by the angle $\theta_\mathbf{a}$. Next, amplitude amplification is used to amplitude-amplify all sequences $\mathbf{a}$ where the reward probability $p_r(\mathbf{a})$ given sequence $\mathbf{a},$ is above a threshold $p_{min}.$

Given $N_{min}$ such sequences (out of $N_{tot} \mathop{:=} |\mathcal{A}|^M $ sequences in total), the overall number of interactions steps multiplies $M$ with the amplitude amplification cost ($O((N_{tot}/N_{min})^{1/2})$), and with phase estimation cost $O(1/p_{min})$.
Overall, we have $O(M (N_{tot}/N_{min})^{1/2}p_{min}^{-1})$ interaction steps. 
The classical agent's interaction cost of the same process is $O(M N_{tot}/
N_{min}).$ 

If the minimal relevant success probability is constant for a family of task environments, then this constitutes a quadratic improvement in finding good action sequences. This approach can also be generalized to a wider class of settings (see the Appendix for details).

In many settings, e.g., robotics, the classical environments do not allow ``oracularization''.
Nonetheless, the presented constructions can be used in model-based learning  \cite{2003_Russel}, where the agent constructs an internal representation of the environment to facilitate better learning through simulation. Then, the ``quantum chip'' can help in speeding-up internal processing, which is the most that can be done in $QC$ settings. A tantalizing exception to this may be nano-scale robots (e.g. intelligent versions of in-situ probes in \cite{2015_Tiersch}) in future quantum experiments, as on these scales the environment is manifestly quantum and exquisite control becomes a possibility. 

\textit{Conclusions.--} In this work we have extended the general agent-environment framework of {artificial intelligence} \cite{2003_Russel} to the quantum domain. Based on this, we have established a schema for quantum improvements in learning, beyond computational complexity. Using this schema, we have given explicit constructions of quantum-enhanced reinforcement learning agents, which outperform their classical counterparts quadratically in terms of learning efficiency, or even exponentially in performance over limited periods. 
This constitutes an important step towards a systematic investigation of the full potential of quantum machine learning, and the first step in the context of reinforcement learning under quantum interaction. 

%WordCountEnd
\begin{acknowledgements}
VD and HJB acknowledge the support by the Austrian Science Fund (FWF) through the SFB FoQuS F 4012, and the Templeton World Charity Foundation grant TWCF0078/AB46. VD thanks Christopher Portmann, Petros Wallden and Peter Wittek for useful discussions which helped in parts of this work.
\end{acknowledgements}

%\newpage

 \begin{widetext}

\section*{Appendix} 
Here we provide further details for the results presented in the main text. The structure of this Appendix follows the structure of the main text, but we provide an outline for the benefit of the reader. In Section \ref{qaep} we introduce the tested quantum agent-environment paradigm, give detailed proofs of the results presented in the main text. In Section \ref{environment-oracles} we give details on the oracular instantiations of task environments, and give constructions for all oracles used later. Following this, in Section \ref{improvements} we give a detailed specification of luck-favoring settings, and give a detailed statement and proof of Theorem 1 from the main text, along with further discussions. Finally, in Section \ref{general} we give further details on how the simpler constructions of the previous sections generalize to broader classes of task environments.
Parts of this Appendix reproduce some of the results of an unpublished, earlier version of this work, available on arXiv \cite{2015_Dunjko}.

\section{Quantum agent-environment paradigm}
\label{qaep}
The basic components of an agent-environment paradigm \cite{2003_Russel} are the set of percepts $(\mathcal{S} = \{ s_i\})$ and the set of actions  $(\mathcal{A} = \{a_j\}$, which are interchangeably issued by the environment, and agent, respectively.
We assume these sets are finite.
A realized interaction up to time step $t$, between the agent and the environment, is a sequence $h_{t} = (s_1, a_2, s_3,s_4, \ldots, s_{t-1},a_{t}), s_i\in \mathcal{S}, a_j \in \mathcal{A}$ of alternating percepts and actions is called \emph{the $t-$step history} of interaction. With $\mathbf{H} = \bigcup_{t\geq 0} \mathbf{H}_t$ we denote the set of all (in principle possible) histories. $\mathbf{H}_t$ denotes the set of all  (in principle possible) histories of length $t$.

The agent and environment are formalized as stochastic maps with memory.

At the $t^{th}$ time-step, and given the elapsed history $h_{t-1}$ the behavior of the agent (the environment) is characterized with the maps
\EQ{
M_{A}^{h_{t-1}}(s\in\mathcal{S} ) \in distr(\mathcal{A});\\
M_{E}^{h_{t-1}}(a\in\mathcal{A} ) \in distr(\mathcal{S}),
}
respectively, 
where $distr(\mathcal{X})$ denotes the set of probability distributions over the set $\mathcal{X}$. The superscripts denote the realized history up to time step $t-1$.  To exemplify, 
the agent outputs, at time step $t$, given percept $s$ and history $h_{t-1}$, the action $a$ which is sampled from the distribution $M_{A}^{h_{t-1}}(s)$.
The agent, and the environment, are defined by the sequences of the maps $\{M_{A}^{h} \}_{h}$, $\{M_{E}^{h}\}_{h},$ indexed over the set of histories.

The agent and the environment may be stochastic.
In this case, with $A\leftrightarrow_t E$ we denote the probability distribution over $t-$step histories, and with $A\leftrightarrow E$ the distribution over all histories $\mathbf{H}$, realized by the agent $A$ and environment $E$. 

The random variable $A\leftrightarrow_t E$ is sometimes referred to as \emph{the interaction} between the agent $A$ and environment $E$. 
We will assume that the interaction begins with the environment outputting the first percept.
To make this formal, we assume that the action and percept spaces contain the empty percept/action element $\epsilon$, which is also the first element of any history, and the first percept of any interaction. Then, given an agent and and environment, the first action output is given with $a \sim \mathcal{M}^{\epsilon}_A(\epsilon)$, followed by the environment's step  $s\sim \mathcal{M}^{\epsilon}_E(a)$. Here, with $x \sim d\in distr(\mathcal{X})$ we mean that the element $x \in \mathcal{X}$ is distributed according to the distribution $d$ over the state space $\mathcal{X}$.  If we require the agent to be on the move first, we will simply assume that the first percept is the empty percept $\epsilon$. 

The definition of interaction is given recursively. The distribution $A \leftrightarrow_2E$ is specified with $P(A \leftrightarrow_2 E = a) = P(\mathcal{M}^{\epsilon}_A(\epsilon) = a),$ where we view the characteristic maps as random variables. The indexing of the interaction starts with 2, as the first move of the environment is defined to be the trivial percept $\epsilon$.

Even length interactions (i.e. ending with the agent's move) are specified with
\EQ{P(A \leftrightarrow_{2t} E = h_{2t}) =  P\left(A \leftrightarrow_{2t-1} E = {\left[h_{2t}\right]}_{-1}\right)P\left(\mathcal{M}^{ {\left[h_{2t}\right]}_{-2} }_A\left({\left[h_{2t}\right]}_{2t-1}\right) =  {\left[h_{2t}\right]}_{2t}\right),
}
where ${\left[h_{2t}\right]}_{-2}$ and ${\left[h_{2t}\right]}_{-1}$ denote the history of length $2t-2$ and and history of length $2t-1$, obtained by dropping the last two elements, and the last element from $h_{2t}$, respectively.  ${\left[h_{2t}\right]}_{2t}$ denotes the last element of the same history. The odd length interactions are defined analogously. 
We do not explicitly model the rewarding step, and assume that the percept space contains a reward-specifying component, so $\bar{\mathcal{S}} =\mathcal{S} \times \Lambda$, where $\Lambda$ is the set of possible rewards, as explained in the main text.

 To extend this setting to the quantum domain, the percepts are represented as orthogonal basis states of the percept Hilbert space  $\mathcal{H}_\mathcal{S}~=~\textup{span} \{ \ket{s} \vert s \in \mathcal{S}  \}$. Analogously, for the action space we have $\mathcal{H}_\mathcal{A} = \textup{span} \{ \ket{a} \vert a \in \mathcal{A}   \}$, also satifying  $\bra{a}{a'}\rangle = \delta_{a,a'}$, where $\delta$ is the Kronecker-delta function.
The agent and the environment have internal memory:  finite-sized registers $R_A$ and $R_E$
which can storine histories, so with Hilbert spaces of the form $\mathcal{H}_\mathcal{A} \otimes\mathcal{H}_\mathcal{S} \otimes \mathcal{H}_\mathcal{A}  \cdots$.

Next, we specify the interface of the agent and the environment - that is parts of the system of the agent (environment) to which they both have access, in contrast to $R_A$ ($R_E$) which are reserved for the agent  (environment) exclusively.
We define the unique common communication register  $R_C$ - the interface - with associated Hilbert space $\mathcal{H}_C$ sufficient to represent both actions and percepts, thus $\mathcal{H}_C~=~\{ \ket{x} | x\in \mathcal{S} \cup \mathcal{A} \}$. The actions and percepts are mutually orthogonal, so $\mathcal{H}_C$ is isomorphic to $\mathcal{H}_\mathcal{S} \oplus \mathcal{H}_\mathcal{A} $.

The agent (environment) is specified by sequences of completely positive trace preserving (CPTP) maps  $\{\mathcal{M}^{A}_i \}_i$ ($\{\mathcal{M}^{E}_i \}_i$) acting on the concatenated registers $R_A R_C$ ($R_C R_E$).
It will sometimes be useful to dilate the maps above to unitary maps, by using appended registers added to $R_A$ and $R_E$ if needed.
An agent-environment interaction is then specified by the sequential application of the maps.

Unless otherwise specified, we will always assume that the initial state of the registers $R_A R_C R_E$ is a fixed product state. Since the initial state does not correlate the agent and the environment, the actual choice of the initial state is not important, as its preparation can be subsumed in the first maps of the agent and environment.

The  classical setting is recovered as follows.
We will call any state, which is a tensor product of percept/action basis states, \emph{a classical state}. Probabilistic mixtures of such states (that is, states whose density operators are convex combinations of the corresponding projectors) are also classical states, and no other states are classical. For completeness we note that classical mixed states are defined relative to a register/system under consideration: for instance, a Bell-pair state of two qubits is not classical, whereas the reduced states of both individual qubit are, as they are equiprobable mixtures of any two orthogonal states.
The definition of classical states is analogous to the standard concept of {computational basis states} in quantum computing. The particular choice of such a basis will, in practice, depend on the particular systems forming the agent and the environment.

\DE
The agent $A$ is \emph{classical}, if for every map $M \in \{ \mathcal{M}^{A}_k \}_k$ acting on $R_AR_C$ the following holds:  if the state of the register $\rho_{R_AR_C}$ is of the product form $\ket{\Psi} = \ket{\psi}_{R_{A}} \ket{s}_{R_{C}},$ where $\ket{\psi}_{R_{A}}$ is a classical state, and $\ket{s}_{R_{C}}$ is a classical percept state, then 
\EQ{
M(\dm{\Psi}) = \sum_{i,j} p_{i,j} \dm{\psi^i}_{R_A} \otimes \dm{a_j}_{R_C},
}
where $\ket{\psi^i}$ are classical states, $\ket{a_i}$ are classical action states and all $p_{i,j}$ are real.
\ED
In other words, the agent $A$ is classical if its maps neither generate entanglement, nor coherent superpositions of classical states, when acting on classical states. {Note that, in the definition above we refer to the defining maps of the agent, not their dilations - the purifying systems (which are not parts of the agents or environments memory) could, naturally, be entangled to both the registers $R_A$ and $R_C$.}
A classical environment is defined analogously.

The setting of a classical agent which interacts with a classical environment constitutes a natural starting point for a sensible definition of a classical interaction. However, for the interaction itself, the internal states of the agent/environment should not matter. Thus we give the following, more general, definition of classical interaction.

\DE
\label{def-class-int}
The interaction between the agent $A$ and the environment $E$ is called \emph{a classical interaction} if at every stage of the interaction, the state of the combined registers can be represented in the form
\EQ{
\rho_{R_A R_C R_E} = \sum_{i} p_i\, \eta^i_{R_A} \otimes \rho^i_{R_C} \otimes \sigma^i_{R_E}
}
where each $\rho^i_{R_C}$ is a convex mixture of classical states, all $\eta^i$ and $\sigma^i$ are unit trace density matrices, and $p_i$ are (real) probabilities.
\ED
Given Definition \ref{def-class-int}, a classical interaction between an agent and the environment does not entail that the agent and the environment are internally in classical states. However, we do prohibit entanglement between the the registers $R_A, R_C$ and $R_E$, and also coherent superpositions in the interface registers.

Next, we consider what, in the context of quantum agent-environment interactions, a proper analog of a history should be.
In the cases where, for instance, the states of the agent, environment and the interface are entangled, there is no straightforward analog of the classical history. Intuitively, since we are dealing with quantum systems, the history should be an observable of the systems.
 More precisely, it should surmount to a sequence of observables, defined for all the time steps of the interaction.
We formalize and and characterize a quantum history more broadly by introducing a third entity - a \emph{tester}. 

The tester is a sequence of CPTP maps $\{ \mathcal{M}^T_k \}$ which act on an external register $R_T$ and the communication register $R_C$ (analogously $R_{I(A)}R_{I(E)}$). Often we may assume that the maps are unitary (by dilating the maps, if needed).
In the \emph{tested} interaction between an agent and the environment, the map of the tester is applied after each map of both the agent and the environment.
The tester is not meant to change the dynamics of the interaction between the agent and the environment, but rather just to `observe' (at least, in the case of a classical interaction). To this end, all the maps of the tester are controlled unitary maps satisfying
\EQ{
U^T_k \ket{x}_{R_C} \otimes \ket{\psi}_{R_T} = \ket{x}_{R_C} \otimes U^{x}_k\ket{\psi} \label{tester2}
} 
where $x \in \mathcal{S} \cup \mathcal{A},$ and $\{ U^{x}_k\}_x$ are arbitrary unitary maps acting on (the subsystems of) $R_T$, for all $k$.

A \emph{classical tester} copies all the states of the $R_C$ register to its own.

To avoid misunderstandings, by copy we mean a unitary map which implements $\ket{x}\ket{\epsilon} \rightarrow \ket{x}\ket{x},$ where $\ket{x}$ is any percept or action basis state. If the duplicate is traced out, the realized map is just a classical basis measurement of the input state.

A tested interaction between the agent and the environment is illustrated in Fig. \ref{fig2}.

\begin{figure}[!h]

\includegraphics[width=0.8\textwidth]{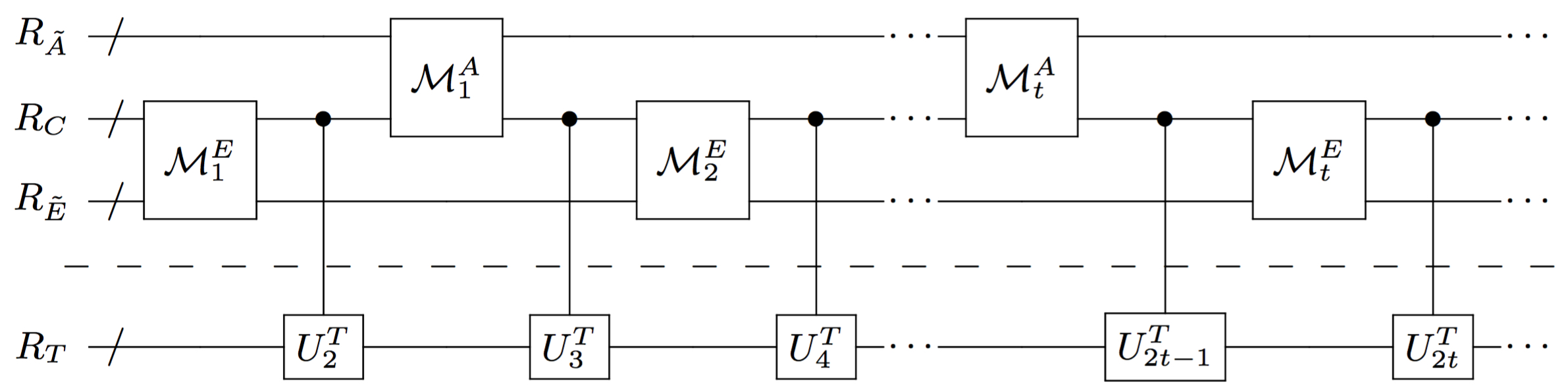}

\caption{\label{fig22}
Tested agent-environment interaction. Note that, in general, each map of the tester $U^T_k$ acts on a fresh subsystem of the register $R_T$, which is outside the control of the agent or the environment. The environments and agents $R_A,$ $R_E$ are subsystems of $R_{\tilde A},$ $R_{\tilde E}$, respectively, along with the purifying registers (possibly) needed for the unitary representation of the maps. The maps of the tester can be assumed be unitary. Each quantum ``wire'' corresponds to an arbitrary number of quantum systems (denoted with the ``$\slash$'' symbol on the wire).
}
\end{figure}

%\begin{figure}[!h]
%\label{fig2}
%\begin{equation}
%\Qcircuit @C=0.7em @R=1.2em {
%\lstick{R_A}&{/} \qw & \multigate{1}{U^A_1}  &\qw& \qw                                 & \qw     & \multigate{1}{U^A_2}& \qw  & \qw                                 & \qw                &\qw&\cdots   &  & \multigate{1}{U^A_t}&\qw & \qw                                           &\qw &\qw       &\cdots    \\
%\lstick{R_C}&{/} \qw &\ghost{U^A_1  }            &\ctrl{3}& \multigate{1}{U^E_1}  & \ctrl{3} &\ghost{U^A_2  }          & \ctrl{3}      & \multigate{1}{U^E_2}  &  \ctrl{3} &\qw&\cdots  &   &\ghost{U^A_t  }      & \ctrl{3}& \multigate{1}{U^E_t}         &\ctrl{3}&\qw &\cdots   &  \\
%\lstick{R_E}&{/} \qw & \qw                                &   \qw   &\ghost{U^E_1  }            &\qw      &  \qw                              & \qw   & \ghost{U^E_2 }          & \qw            &\qw &\cdots      && \qw                             &  \qw & \ghost{U^E_t  }                   & \qw&\qw & \cdots &\\
%&\\
%\lstick{R_T}&{/} \qw & \qw                                &\gate{U^T_1}& \qw         &\gate{U^T_2} &  \qw                                 &\gate{U^T_3}   & \qw         & \gate{U^T_4}    &\qw          & \cdots      && \qw &   \gate{U^{T}_{2t-1}}     & \qw &\gate{U^T_{2t}} &\qw & \cdots &
%}
%\end{equation}
%\caption{Tested agent-environment interaction for the communication model.}
%\end{figure}

The classical history is then recovered by the sequence of states of the register $R_T$, relative to the classical tester. A general quantum history is given by considering the state of the register $R_T$ without placing any (additional) restrictions on the maps of the tester, aside from the fact that we require them to be of the `classically controlled' form given in Eq. (\ref{tester2}). 

%Hence all the statements we will be making will hold for both the communication and the embodied model.

A few remarks are in order. In the case of stochastic classical agents (environments), the agent (environment) will, at each time step, output a particular action (percept) with some probability. In the quantum model, this will be represented by the agent outputting a convex mixture of action states, specified by the corresponding probabilities of the particular action. Thus, in the setting of a classical tester $T_c$, the state of the register $R_T$, at time-step $t$, can be expressed, in terms of the classical agent-environment interactions, as follows:
\EQ{
\rho^t_{T_c}(A,E) = \sum_{h_t \in \mathbf{H}_t} P(A \leftrightarrow_t E = h_t) \dm{h_t},\label{qt}
}

The above is exactly the classical history, defined previously, represented in the standard quantum formalism. 
The quantum history state $\rho^t_T(A,E)$ will, in general, attain significantly different forms for different testers, and we will refer to it as \emph{the quantum history between agent and the environment, relative to the tester $T$}. In the quantum interaction case, a figure of merit for learning will be a function of a quantum history of the interaction.

The presence of a classical tester changes nothing in the case of classical agents and environments. It is also not problematic for agents and environments which  only have a classical interaction, which is slightly more general:

\LE
\label{LE-equiv}
For any agent $A$ and environment $E$,
$A$ and $E$ have a classical interaction if and only if the (reduced) state of the three registers of $R_{A}R_C R_E$ is the same in the presence and absence of a classical tester.
\EL
\proof

$(\Longrightarrow)$ If $A$ and $E$ have a classical interaction, by definition, at each stage of interaction, the state of the three registers $R_{A} R_C R_E$ is of the form 
\EQ{
\rho_{R_A R_C R_E} = \sum_{i} p_i\, \eta^i_{R_A} \otimes \rho^i_{R_C} \otimes \sigma^i_{R_E}.
} 
Moreover,  we have that ${\displaystyle \rho^i = \sum_{j \in \mathcal{S} \cup \mathcal{A}} q^i_j \dm{j}},$ for $q^i_j \in \mathbbmss{R}^+\cup \{ 0\}$ . Applying the classical tester yields the following state of $R_A R_C R_E R_T$:

\EQ{
\rho_{R_A R_C R_ER_T} =  \sum_{j \in \mathcal{S} \cup \mathcal{A}} \sum_{i}   p_i q^i_j\, \eta^i_{R_A} \otimes \dm{j}_{R_C} \otimes \sigma^i_{R_E} \otimes \dm{j}_{R_T},
} 
where we have, for clarity, commuted the sums over $i$ and over $j$.
It is now obvious that tracing out $R_T$ just recovers $\rho_{R_A R_C R_E},$ so this implication holds. \\
\vspace{0.5cm}

\noindent $(\Longleftarrow)$  We prove this direction by induction over interaction steps. Suppose the claim holds up to step $t-1,$ so at that time-step, the state of the three registers is 
 \EQ{
\rho^{t-1}_{R_A R_C R_E} = \sum_{i} p_i\, \eta^i_{R_A} \otimes \rho^i_{R_C} \otimes \sigma^i_{R_E}.
} 
Next, it is either the environmental or the agent's move. We will assume it is the agent's move, and the claim for the case of the environment's move can be shown analogously.
The agent's map only sees registers $R_AR_C$ so we can write the state of the subsequent step as 
 \EQ{
\rho^{t}_{R_A R_C R_E} = \sum_{i} p_i\, \eta^i_{R_AR_C} \otimes \sigma^i_{R_E}.
} 
Now, each $ \eta^i_{R_AR_C}$ can be written as a convex combination of pure states:
\EQ{
\eta^i_{R_AR_C} = \sum_{j}{q^i}'_j {\dm{\psi_{i,j}}}_{R_AR_C},
} and each pure component $\ket{\psi_{i,j}}$ can be decomposed w.r.t. a separable basis:
\EQ{
\ket{\psi_{i,j}} = \sum_{k,l} \alpha_{k,l}^{i,j} \ket{\phi_{k}} \otimes \ket{x_{l}},
}
where $\ket{\phi_{k}}$ are classical states and $\ket{x_{l}}$ is a percept or an action state.
Putting it all together we have:
\EQ{
\eta^i_{R_AR_C} = \sum_{j}{q'}^i_j    \sum_{k,l,k',l'}      \alpha_{k,l}^{i,j}{\alpha_{k',l'}^{i,j}}^{\ast}\ket{\phi_k}\bra{\phi_{k'}}_{R_A} \otimes \ket{x_{l}}\bra{x_{l'}}_{R_C}
}
Copying the $R_C$ register (w.r.t. the classical basis), and then tracing out the copy system, reduces to eliminating all cross terms $ \ket{x_{l}}\bra{x_{l'}},$ where $l \not=l'.$ In other words, the following must hold, in order for the state to be invariant under classical testing:
 \EQ{
\sum_{i} p_i\,  \sum_{j}{q'}^i_j    \sum_{k,l,k',l'}      \alpha_{k,l}^{i,j}{\alpha_{k',l'}^{i,j}}^{\ast}\ket{\phi_k}\bra{\phi_{k'}}_{R_A} \otimes \ket{x_{l}}\bra{x_{l'}}_{R_C} =\\  \sum_{i} p_i\, \sum_{j}{q'}^i_j    \sum_{k,l,k',l'}      \alpha_{k,l}^{i,j}{\alpha_{k',l'}^{i,j}}^{\ast} \delta_{l,l'}\ket{\phi_k}\bra{\phi_{k'}}_{R_A} \otimes \ket{x_{l}}\bra{x_{l'}}_{R_C},
 }
where $\delta_{l,l'}$ is the Kronecker-delta.
So,
\EQ{
\rho^{t}_{R_A R_C R_E} =  \sum_{i}  \sum_{l} p_i\, \sum_{j}  \sum_{k,k'}  {q'}^i_j     \alpha_{k,l}^{i,j}{\alpha_{k',l}^{i,j}}^{\ast} \ket{\phi_k}\bra{\phi_{k'}}_{R_A} \otimes \ket{x_{l}}\bra{x_{l}}_{R_C} \otimes \sigma^i_{R_E},
}
 and by defining $\eta'_{l,i}=\sum_{j}  \sum_{k,k'}  {q'}^i_j     \alpha_{k,l}^{i,j}{\alpha_{k',l}^{i,j}}^{\ast}   \ket{\phi_k}\bra{\phi_{k'}} $ we get 
 \EQ{
 \rho^{t}_{R_A R_C R_E} =  \sum_{i} p_i\,   \sum_{l} {\eta'_{l,i}}_{R_A} \otimes \ket{x_{l}}\bra{x_{l}}_{R_C} \otimes \sigma^i_{R_E} = \sum_{i,l} p_i \,  {\eta_{l,i}}_{R_A} \otimes \dm{x_l}_{R_C} \otimes \sigma^i_{R_E}.
 }

The expression above is of the desired form, as soon as the sum is re-written with one running index.
Analogously, we obtain the claim for the environment's first move. This shows the step of the inductive proof, as the invariance under classical testing guarantees we always go from desired form states to desired form states.
To finish the inductive proof, we must establish the base of the induction. However, as we have clarified before, we assume that the initial state of the registers of the agent and environment is in product form, so this is trivial. \qed
\\

Examples of such `internally quantum' agents which interact classically is, for instance, a standard (classical input-classical output) quantum computer, where the environment would be the users.
Such (internally only) quantum agents and environments which have a classical interaction cannot offer different behaviors, compared to classical agents interacting with classical environments, relative to any tester:
\LE
\label{LE-clas-int}
For any agent $A$ and environment $E$, which have classical interaction (when untested), there exists 
a classical agent $A^c$ and a classical environment $E^c$, such that $\rho^t_T(A,E) = \rho^t_T(A^c,E^c),$ for any tester $T,$ and any history length $t$.
\EL
\proof

This lemma essentially follows from the classical simulability of quantum mechanics. In particular, we can consider the classical agent $A^C$ and the classical environment $E^C$, which, internally, instead of storing quantum states, store the classical descriptions of the same quantum states: if the joint system, at time step $t-1$ of the registers $R_AR_CR_E$ is:
\EQ{
\rho_{R_AR_CR_E} = \sum_{i} p_i\, \eta^i_{R_A} \otimes \rho^i_{R_C} \otimes \sigma^i_{R_E}
 \label{quant}}
the corresponding state generated by $A^C$ and $E^C$ would be

\EQ{
\rho_{R_{A^C}R_CR_{E^C}} = \sum_{i} p_i\, [\eta^i]_{R_A} \otimes \rho^i_{R_C} \otimes [\sigma^i]_{R_E},
 \label{class}}
where with $[\rho]$ we denote the numerical matrix of the quantum state $\rho$.
To clarify, the classical description $[\rho]$ of the quantum state $\rho$ is also a quantum state. However, note that it is always also a classical state as $[\rho]$ and $[\rho']$ are orthogonal whenever $\rho \not= \rho'$ (we can perfectly distinguish two distinct \emph{matrices}, regardless of the fact that they may \emph{represent} non-orthogonal \emph{quantum states}). This may imply an exponential blow up in the number of registers needed, (and in the computation time), but this is irrelevant in the synchronous model of agent-environment interaction.

The transition to state $t$ is achieved by applying a map of the agent, or environment. Suppose it is the agent's move, as the argument will be analogous for the environmental move case.
At this point, the agent will apply a quantum map $\mathcal{M}$ to its system and the register $R_C,$ which maps 
\EQ{\eta^i_{R_A} \otimes \rho^i_{R_C} \stackrel{\mathcal{M}}{\longrightarrow} \sum_j q_j\   {\eta'}^j_{R_A} \otimes {\rho'}^j_{R_C},}
where the particular structure is ensured by the assumption the interaction is classical.
The classical agent can then be defined to apply a corresponding map mapping 

\EQ{[\eta^i]_{R_A} \otimes \rho^i_{R_C} \stackrel{[\mathcal{M}]}{\longrightarrow} \sum_j q_j\   [{\eta'}^j]_{R_A} \otimes {\rho'}^j_{R_C},
}
which is possible because the state $\rho^i$ is already a classical state, as the interaction is classical.
This establishes an inductive step. The basis of the induction also holds, provided that the initial state of the registers $R_AR_CR_E$ is a classical state, which, as we have clarified, we assume to be the case.
By inspecting equations $(\ref{quant})$ and  $(\ref{class})$ specifying the structure of the states of the registers realized by $A$ and $E$ and the classical counterparts $A^C$ and $E^C$ it is clear that the quantum histories generated by the two will be the same for all testers. \qed

In particular, this also implies that the two learning settings will have identical figures of merit, relative to any figure of merit which depends only on the history.

%Any figure of merit which is a function of the history alone can be achieved by utilizing quantum mechanics, if the interaction is classical.

Similarly, in the presence of a classical tester, quantum improvements are also not possible:

\LE
\label{LE-clasTest}
Let $A$ and $E$ be any agent and any environment over compatible percept/action spaces.
Then there exists a classical agent $A^c$ and a classical environment $E^c$, such that $\rho^t_{T_c}(A,E) = \rho^t_{T_c}(A^c,E^c),$ for the classical tester $T_c,$ and any history length $t$.
\EL
\proof
Adding an additional classical tester (instead of just one) still generates the same quantum history within the original classical tester. However, tracing out the register of the additional tester reduces the interaction of $A$ and $E$ to a classical interaction, as all non-classical terms (off-diagonal components in the states of $R_C$) are removed by the trace-out. But then by Lemma \ref{LE-clas-int}, the same quantum interaction generated by a classical tester can be achieved by a classical agent and a classical environment. Note that we cannot use the same argument for other testers - adding a second classical tester may change the quantum history generated by another type of tester. \qed
 
The results above should not be particularly surprising -- classical interactions simply lack the capacity for sufficiently subtle control to allow for any quantum effects (including speed-ups) almost by definition.
Thus, we consider other types of testers to achieve improvements.
 In this work, we will focus on the \emph{sporadic} tester, which allows for periods of untested, fully coherent interaction, followed by classically tested interaction. While this is still a restricted setting, maintaining the tester fully classical at periods will allow for a straightforward comparison between quantum and fully classical agents. 
 
% We note that, alternatively, the framework of quantum combs \cite{2008_Chiribella} could be used to model the environment-agent-tester interaction, and indeed, the generality of that approach may become beneficial in settings with multiple agents. We also note that the authors of the last reference and collaborators in \cite{2013_Chiribella} use the terms `quantum tester' and refer to `histories of classical communication', albeit in a cryptographic context, which are remotely related to the ideas we have presented. 
 
In the remainder of this Appendix, we will use the term \emph{fully classical agent} to refer to an agent which is classical, but also forces the interaction (for any environment) to be classical. Here, forcing implies that, within the model, the agent always de-phases the register $R_C$ (equivalently, registers $R_{I(A)}R_{I(E)} $), by e.g. classical basis measurements, whose outcomes are discarded. 
Since, for the purposes of this work, we are interested in quantum enhancements of classical learning agents, in the next section we consider what kinds of quantum extensions classically specified environments in principle allow.

\subsubsection{The generic performance of a quantum-enhanced agent}
\label{generic-performance}

Suppose we are faced with a classical learning scenario, with a fully classical agent $A$ and an environment $E$, which is, \emph{a priori} unknown.
We would then like to asses the properties of the interaction of a quantum agent $A^q$, for the purposes of comparison,  with the \emph{same} environment $E,$ which can now be accessed via a quantum (not classical, in the sense of the definition we have in the previous section) interaction.
The question then is, in general, when can we consider two environments $E$ and $E'$ to be `the same'. There are a few natural answers.
The strongest notion of sameness would demand that two environments are equal, if they are specified by the same sequence of CPTP maps. A weaker notion of sameness is \emph{equivalence relative to the tester $T$}: $E$ and $E'$ are equal relative to $T$ if the quantum histories of $E$ and $E'$, relative to the tester $T$, are the same for any agent. If two environments satisfy the stronger notion of sameness, then they are equal relative to all testers. Note that all environments are equal relative to trivial testers, which apply the same map irrespective of the states in the communication register.
%A similar statement can be made in reverse.
However, since we are adopting the approach of extending classical learning scenarios to quantum for the purpose of comparison, we are interested in the following definition:
\DE
Two environments $E$ and $E'$ are the equal \emph{in the classical sense}, denoted $E=_c E'$ if they are the equal relative to the classical tester.
\ED
The above is equivalent to saying that $E$ and $E'$ are the same in the sense of realizing identical classical distributions over histories for any fully classical agent. The definitions above could also be relaxed to approximate equalities (within some distance) by relaxing the equalities on the quantum histories (using e.g. an approximate equality on states induced by the trace distance).

It is easy to see that the equality in the classical sense is an equivalence relation on environments.
For each environment $E$ we can then identify the classical equivalence class $\mathcal{E}_c(E) = \{E' | E =_c E' \}.$ All the elements of the class $\mathcal{E}_c(E)$ share the property that the classical maps they realize (in the sense of the classical definition of agent-environment interaction), in a classical interaction, are equal for all environments in the class.
This sequence of classical maps (i.e. this classical environment) we will call \emph{the classical specification} of the class $\mathcal{E}_c(E)$. Then we will also say an environment $E$ is \emph{only classically specified} if only its classical specification is known. 
Recall, in fully classical learning, classical specification is all there even is.
The next simple lemma states that if only the classical specification of an environment is known, no quantum enhancement can be generically guaranteed. 
\LE
\label{lem-extens}
Let $\mathcal{E}_c(E)$ be the classical equivalence class for some environment $E$. Then there exits a quantum environment $E^q \in \mathcal{E}_c(E)$ which prohibits any quantum improvement -- that is, any possible quantum history (relative to any tester) can be realized with a fully classical agent and this environment $E^q.$
\EL
\proof Take any environment $E' \in \mathcal{E}_c(E),$ and sandwich every CPTP map which specifies the environment $E'$ with a classical basis measurement of the register $R_C$ (equivalently, $R_{I(A)}R_{I(A)}$).
This is a new environment, $E^q,$ it is clearly in $\mathcal{E}_c(E)$, but it also forces a classical interaction. Then by Lemma \ref{LE-clas-int}, no quantum advantage is possible in this environment for any agent. \qed

The lemma above should be clear. With the permission of a bit of poetic license, it asserts that just putting on our ``quantum eyeglasses", that is, acknowledging that any real system is a  quantum system with quantum degrees of freedom, does not turn, for instance, a classical computer into a quantum computer. Even with fully coherent quantum input, most devices (or environments) will have decoherence processes which prevent any true quantum dynamics on any useful scale.
While this observation is straightforward, it is nonetheless relevant for our case. In what follows, we will begin by specifying an interaction between a classical environment and a quantum agent. Then, we will ask whether the quantum agent could do better, if the environment can be accessed as a quantum system, and the agent is free to exploit quantum coherence. Lemma \ref{lem-extens} then asserts that the answer may be a trivial no, unless further assumptions are made on \emph{how} the environment extends to a full quantum system. We acknowledge that, from a physics point of view, it would be more natural to consider this problem in reverse. Any physical system is fundamentally quantum, and one can consider classical limits of the quantum system, rather than `quantum extensions' of an otherwise classical systems. However, in the spirit of the mainstream approaches to artificial intelligence, systems, and task environments are usually assumed to be classical, both in the computational tradition and in robotics.
From this perspective, since we start from such classical problem, it makes sense to talk about quantum extensions, that is, quantum systems which are compatible with the given classical limit.

The question of what are useful quantum extensions of classically specified \emph{functions} is also vital in the case of quantum computation with the aid of a quantum oracle.

\section{Basic oracular instantiations of task environments}
\label{environment-oracles}
\subsection{Oracles for deterministic environments}
First, we consider the special case of deterministic environments $E$ where the state of the environment is re-set after exactly $M$ steps. In this case, all dependence of the environmental memory on the actions of the agent is lost after each block of $M$ steps.
In other words, we can express the environment as a reversible map acting on $M$ moves simultaneously.
For simplicity we will use bold-face fonts to denote a sequence of indexed symbols, so e.g.  $\ve{a}\mathop{:}= (a_1, \ldots, a_M)$. Such an $M$-block instantiation is given with the following expression:
\EQ{
\ket{\ve{a}}_{\textup{I}} \ket{\ve{y}}_{\textup{II}} \stackrel{U_E}{\longrightarrow} \ket{\ve{a}}_{\textup{I}} \ket{\ve{y} \oplus \ve{\bar{s}}}_{\textup{II}}\ ,
}
where $\ve{y} = (y_i)_i$ and $y_i$ is an element of the percept set $\mathcal{S}$, and $\ve{y} \oplus \ve{\bar{s}} = (  y_i \oplus \bar{s}_i    )_i,$ and $\oplus$ denotes a group operation on the set $\mathcal{S},$ e.g. the modulo-$|\mathcal{S}|$ addition on the set of indices specifying the elements of the $\mathcal{S}$. Here, $\bar{s}$ is the sequence of percepts that the environment will deterministically output if the agent outputs the sequence of actions $\ve{a}$. 
The group operation can be chosen such that the induced group is of order 2 (except for the identity element) - if $|S| = 2^k,$ for some $k \in \mathbbmss{N}$ then the indices can be represented as binary string, and the addition is bitwise mod-2 addition.
Then, $U_E$ is also self-inverse (hence also Hermitian).

In the case only one, last, percept carries a binary reward status, then $O_E = U_E Z_{\Lambda} U_E $ can easily be turned into the phase-flip oracle we used in the main text, where $Z_{\Lambda}$ induces a global (-1) phase, whenever the reward status of any percept is rewarding.  This is the standard ``phase kick-back'' method.
Explicitly, the oracle is given by the mapping $\psi \rightarrow tr_{\textup{II}}\left[  O_E\, \left(\psi_{\textup{I}} \otimes \dm{{0}}_{\textup{II}} \right) O^\dagger_E  \right],$ where $\ket{0}$ is an arbitrary state, for any quantum state $\psi$.

\subsection{Oracle for stochastic settings}

A stochastic environment which deterministically re-sets after $M$ steps can be represented by the following CPTP mapping:
\EQ{
\dm{\ve{a}} \otimes \dm{0} \stackrel{\mathcal{E}_E}{\longrightarrow} \dm{\ve{a}} \otimes \sum_{\ve{\bar{s}}} P(\ve{\bar{s}} \vert {\ve{a}} ) \dm{\ve{\bar{s}}},
} 
where $P(\ve{\bar{s}} \vert {\ve{a}} )$ is the probability of the environment outputting the percept sequence $\ve{\bar{s}} =(\bar{s}_i )_i,$ when the agent performs the sequence of actions $ {\ve{a}}$. Recall, in the standard classical setting, the exchange of percept-actions is interchangeable, but we can nonetheless represent this interactive process as the $M-$ block map above.
For any such stochastic map there exists a purifying unitary map which realizes the same dynamics on the reduced system, and this map can be easily constructed. If we represent the classical stochastic process as an invertible map which is also acting on a register containing a random bit-string, the purification of this process would be the corresponding unitary mapping, acting on a purification of the random bit string. 
In this case, the conditional mixed state of percepts $ \pi^{\ve{a}}= \sum_{\ve{\bar{s}}} P(\ve{\bar{s}} \vert {\ve{a}} ) \dm{\ve{\bar{s}}}$ is purified by a part of the environment, and the purification is (up to local unitaries) equivalent to the state
\EQ{
 \ket{\pi^{\ve{a}}}= \sum_{\ve{\bar{s}}} \sqrt{P(\ve{\bar{s}} \vert {\ve{a}} ) } \ket{\ve{\bar{s}}} \ket{\ve{\bar{s}}} .
}
Then, if this purification, instead of the mixed state is returned to the agent, for the simplest case of single reward,  the environment can be represented by a unitary mapping performing
\EQ{
\ket{\ve{a}}_{\textup{I}} \otimes \ket{0}_{\textup{II}}\otimes \ket{0}_{\textup{III}} \stackrel{S_E}{\longrightarrow} \ket{\ve{a}}_{\textup{I}} \otimes \left( \sqrt{p_{\lambda=0}}  \ket{\pi_{\lambda=0}^{\ve{a}}}_\textup{II}\ket{0}_\textup{III} + \sqrt{p_{\lambda=1}}  \ket{\pi_{\lambda=1}^{\ve{a}}}_\textup{II} \ket{1}_\textup{III} \right), \label{stoch1}
} 
where the register $\textup{III}$ represents the rewarding status, and everything is implemented reversibly.
Unfortunately, the state in Eq. (\ref{stoch1}) is not of the suitable form to realize the stochastic oracle assumed in the main text, as the percept register  $\textup{II}$ is, in general, entangled to the reward register $\textup{III}$. This can be resolved in special cases only, and to achieve the generic form of the oracle, we will introduce further assumptions on how the environment is constructed (i.e. further specify its quantum extension).
In particular, note that the aspect of the environment which determines the raw percept sequences can be realized by a controlled map:
\EQ{
\ket{\ve{a}}_{\textup{I}} \otimes \ket{0}_{\textup{II}} \rightarrow \ket{\ve{a}}_{\textup{I}} \otimes U^{\ve{a}}\ket{0}_{\textup{II}},
}
where $U^{\ve{a}}\ket{0}= \ket{\pi^{\ve{a}}}$. If we assume that the set of maps $\{ U^{\ve{a}} \}_{\ve{a}}$ has a known common $+1$ eigenstate $\ket{\phi},$ then the construction of the stochastic oracle can always be achieved. This assumption is justified, e. g. if the Hilbert space of the percept space is increased by one dimension to include the state $\ket{\phi},$ defined to be orthogonal to the basic Hilbert space $\mathcal{H}_{\mathcal{S}},$ and the unitaries $\{ U^{\ve{a}} \}_{\ve{a}}$ are extended such that they act as the identity on that additional one-dimensional subspace.
We note that this is essentially equivalent to assuming that the environment can be forced to just output the reward status, rather than the entire sequence. In the classical case this is a trivial assumption as we can always just ignore the percepts, but in the quantum case, this would correspond to a trace-out operation which would break the desired superposition of the reward status, so we must be more careful.

Assuming, however, that such a common $+1$ eigenstate $\ket{\phi}$ is known, we can obtain the mapping 
\EQ{
\ket{\ve{a}}_{\textup{I}} \otimes \ket{\phi}_{\textup{II}} \otimes \ket{0}_{\textup{III}} \stackrel{S_E}{\longrightarrow} \ket{\ve{a}}_{\textup{I}} \otimes \left( \sqrt{p_{\lambda=0}}  \ket{\phi}_\textup{II}\ket{0}_\textup{III} + \sqrt{p_{\lambda=1}}  \ket{\phi}_\textup{II} \ket{1}_\textup{III} \right) =
\ket{\ve{a}}_{\textup{I}} \otimes  \ket{\phi}_\textup{II} \otimes\left( \sqrt{p_{\lambda=0}} \ket{0}+ \sqrt{p_{\lambda=1}} , \ket{1}\right)_\textup{III} 
}
which is isometrically equivalent to the stochastic oracle defined in the main text, and it can be realized in a self-inverse fashion.

\subsection{Oracle for multiple rewards settings}
\label{counting}
Counting oracles can be constructed in a similar manner: starting with the reversible, self-inverse instantiation of the environment $U_E,$ we append a count register and apply the operation $C^{\Lambda}$ which counts the total reward of the sequence:
\EQ{
 \ket{\ve{\bar{s}}}_{\textup{II}} \otimes \ket{y}_{\textup{III}} \stackrel{C^\Lambda}{\longrightarrow}  \ket{\ve{\bar{s}}}_{\textup{II}} \otimes \ket{\Sigma \lambda \oplus y}_{\textup{III}}, }
where $\Sigma \lambda$ denotes the total of the rewards appearing in the sequence $\ve{\bar{s}} = ( \bar{s}_i    )_i. $ Overall we obtain the mapping
\EQ{
\ket{\ve{a}}_{\textup{I}} \ket{\ve{\bar{s}}}_{\textup{II}} \ket{y}_{\textup{III}} \stackrel{U_{Count}}{\longrightarrow}  
\ket{\ve{a}}_{\textup{I}} \ket{\ve{\bar{s}}}_{\textup{II}}  \ket{\Sigma \lambda \oplus y}_{\textup{III}},
}
which, as we have clarified earlier, can be Hermitian, hence self-inverse.
Using phase-kick back again, we can achieve a reflection operator about the subspace of all sequences $\ket{\ve{a}}_{\textup{I}},$ satisfying the property that the total reward is above a certain chosen value.
 
Note, the construction remains unchanged even if the rewarding set contains rewards of different magnitudes, with the understanding that the reward-carrying register is large enough to represent any sums which may appear.

\section{Quantum improvements and luck-favoring settings}
Here we prove the main theorem from the main text in full detail. First, we give a detailed definition of luck-favoring settings.
\DE
Let $A$ be a learning model/agent and $E$ a legitimate (with matching percept-action structure) environment of A.

Let $Rate(\cdot)$ denote a learning-related figure of merit, defined on histories and extended to distributions over histories by convex-linearity (e.g. the average reward of a history per time-step).

Then we say that the pair $(A, E)$ is \emph{monotonically luck-favoring} for histories $h^{E}_t$ and $\tilde{h}^{E}_t$ relative to the merit function $Rate(\cdot)$ if 
\EQ{
Rate(h^E_n) \geq Rate(\tilde{h}^E_t) \Rightarrow Rate(A(h^E_t) \leftrightarrow E(h^E_t)) \geq Rate(A(\tilde{h}^E_t) \leftrightarrow E(\tilde{h}^E_t)), \label{main-eq-luck}}

where $h^{E}_t$ and $\tilde{h}^{E}_t$ denote two (classical) histories of length $t$ that could have been generated by an interaction of $A$ with $E$, thus:
 \EQ{P(A\leftrightarrow_{t} E = h^{E}_t) \neq 0\ & \textup{and}\nonumber\\ P(A\leftrightarrow_{t} E = \tilde{h}^{E}_t) \neq 0.& \label{assumptionNotZero}}
If Eq. (\ref{main-eq-luck}) holds for any two histories, then we say $(A,E)$ is monotonically luck favoring for all histories.

More specifically, we may be interested in the behavior for specified numbers of interactions $t,t'$. Then we say that
$(A, E)$ is \emph{monotonically luck-favoring} for the merit function $Rate(\cdot),$ with a t-step preparation $(h^{E}_t, \tilde{h}^{E}_t)$, followed by $t'$ step evaluation if
\EQ{
Rate(h^E_t) \geq Rate(\tilde{h}^E_t) \Rightarrow Rate(A(h^E_{t}) \leftrightarrow_{t'} E(h^E_{t})) \geq Rate(A(\tilde{h}^E_t) \leftrightarrow_{t'} E(\tilde{h}^E_t)) \label{main-eq-luck2}
}
%\blue{We will say  the pair $(A, \mathcal{S}_E)$ is \emph{strongly} monotonically luck-favoring if the implication above holds for histories of differing lengths $h^E_n$ and $\tilde{h}^E_{n'}$ with $n \not= n'$.}
\ED

A few comments regarding the definition above are in order. First, $A(h_t)$ and $E(h_t)$ denote agents/environments which have undergone history $h_t$. Technically speaking $A(h_t)$ can be thought of as a different agent from $A,$ which we can write as $A(\epsilon),$ that is, the agent $A$ which has undergone the trivial history $\epsilon$, and the same holds for the environment.
Nonetheless, the `raw' agent $A$ and an `experienced' agent $A(h_t)$ have the same percept-action structure.
The assumptions around Eq. (\ref{assumptionNotZero}) guarantee that the given histories could have been generated by the interaction of the agent and environment, and this technical assumption will be relevant in the construction of the quantum enhanced agent.

The basic idea behind this construction is to first use quantum access to the environment to find a rewarding sequence $h_r$ faster than a classical agent could. Following this, using this sequence, an internal simulation of a classical agent $A$ can be trained to produce exactly that sequence - in other words, the interaction between the environment and the agent is simulated iteratively, until, by chance alone, the agent $A$ `gets lucky' and produces the winning sequence in the first try.
In luck-favoring settings, such agent will outperform an agent which was not lucky later on, by definition.
However, there are a few details which me need to iron out first.

To present our result regarding the speed-up in learning in a clean form, we shall place additional assumptions on the environment  $E$ aside from it being deterministic, single-win and fixed-time. Additionally, we will assume that there is only one winning action path of the length $M,$ where $M$ is also the allotted fixed time (in this case, there is only one winning history of length $M$).
{Recall that, in the case of the maze environment we have described, this implies that the agent traversing the maze is always returned to the $Start$ vertex after exactly $M$ steps.}

 Let $n$ be the size of the action space, thus $n = |\mathcal{A}|.$
Then, the classical agent will require, on average, $O(n^M \times M)$ interaction steps with the environment, before encountering the winning path (note that each `testing' of a particular sequence of action costs $M$ interaction steps).
The quantum agent, given access to the oracular instantiation $E^q_{oracle},$ can achieve the same in expected time $O(\sqrt{n^M} \times M)$, using the (randomized) Grover's algorithm \cite{1996_Grover, 1998_Boyer}.
This constitutes a quadratic improvement in the exploration phase of learning, and what remains to be seen is how to embed this into the complete learning package.

Both the classical and the quantum agent we will now construct are situated in the same controllable environment, namely, $E^q_{control}$. The classical agent $A$ has nothing to gain from quantum oracular access and its access to this environment is only via its classical instantiation $E$.

For the classical agent $A$ (and its underlying learning model) we will next define a corresponding quantum agent $A^q$. Following the precise specification, we will briefly comment on the basic ideas behind the construction.

Since $A$ is fixed and known, we will assume $A^q$ has black-box access to (a simulation of) the agent $A$. In particular, $A^q$ can, internally, feed the simulation of $A$ with any sequence of percepts, and observe the output actions. Moreover, it can always reset the simulation to the initial state as defined for the agent $A$.
Since we are constructing $A^q$ given a classical agent $A,$ we in principle have access to every aspect of $A$ (its program, realization and specification of each characteristic map), but for our purposes, black-box access, and the capacity to reset will suffice.
As a technical assumption, we will assume that the agent $A$ has a non-zero probability of hitting the rewarding sequence of actions, starting from its initial configuration. 
We give a formal specification of the quantum agent $A^q$ next, followed by an explanation of the purpose of each of the steps.

\begin{enumerate}
\item \label{prep1} For the first $t' = k \times\sqrt{n^M} \times{M} $ time steps, $A^q$ engages in a Grover-type search for the awarded sequence of actions, interacting with $E^q_{oracle}$. Recall that each access to the oracle incurs $M$ interaction steps, thus we total $ k \times\sqrt{n^M}$ oracular queries, where $k$ is an integer we specify later.
The agent $A^q$ succeeds in finding the winning sequence $(a_1, \ldots, a_{M}),$ except with probability in $O(\exp(-k))$, since the fraction of winning versus the total number of sequences is $n^{-M}$.
Recall, Grover's algorithm may fail to produce the target element, but this occurs with probability less than 1/2. Iterating the algorithm $k$ times ensures that a failure can occur at most with an exponentially decaying probability in $k$, as stated.

\item \label{prep2} For the next $M$ time-steps, the agent $A^q$ engages the non-oracularized quantum extension $E^q$ of $E$ (or $E$ itself), outputs the (classical) actions $a_1, \ldots, a_{M},$ sequentially and collects the unique corresponding outputs $s_2, \ldots s_{M+1}$ from the environment (by convention, we set the first percept of the environment to be the empty percept $\epsilon$). The entire rewarding history is \EQ{h_{win}= (s_1, a_1, s_2, \ldots, a_{M}, s_{M+1}).} This step is necessary as the oracular access, by construction, does not provide the perceptual responses of the environment.

\item \label{training} Between the time steps $t = t'+M$ and  $t = t'+M+1$, $A^q$ `trains' a simulation of $A$ internally:
It runs a simulated interaction with $A,$ by giving percepts $s_1, \ldots, s_{M+1}$. It aborts and restarts the procedure (with a reset of the simulation of the agent $A$)  until $A$ responds with $(a_1, \ldots, a_{M})$. 
By the technical assumption we mentioned earlier, the expected time of this event is finite.  
The training procedure itself, for the $M$ time steps, is repeated sequentially, until the same winning sequence of actions of the simulated agent $A$ is produced $1+k \sqrt{n^M}$ times, again contiguously. 
Since one sequence can be attained in finite time, so can any finite repetition of the sequence.
This technicality we further explain later.

During this time the agent $A^q$ does not communicate to the environment, and uses up no interaction rounds.
\item \label{trained} Internally, $A^q$ has a simulation of the agent $A(h_{tot})$, with
\EQ{
h_{tot}=\underbrace{h_{win} \circ \cdots\circ h_{win}}_{(1+k\times\sqrt{n^M})\ times}
}
and $\circ$ denotes the (string-wise) concatenation of histories. From this point on, $A^q$ simply forwards the percepts and rewards between the simulation and the environment. 
\end{enumerate}

To talk about learning properties of the defined quantum agent we need to specify the tester.
To optimize our result, we select the sporadic classical tester $T_S$ which is defined as follows:

For the first $t = k \times\sqrt{n^M}\times{M}+M$ time-steps, with $k \in \mathbbmss{N},$ the sporadic tester allows for completely untested interaction. After the $t$ steps, the tester $T_S$ behaves as the classical tester.

This finishes our specification of the quantum-enhanced learning setting, and it is illustrated in Fig. \ref{construction}.

\begin{figure}
\centering

\fbox{\includegraphics[width=0.8\textwidth]{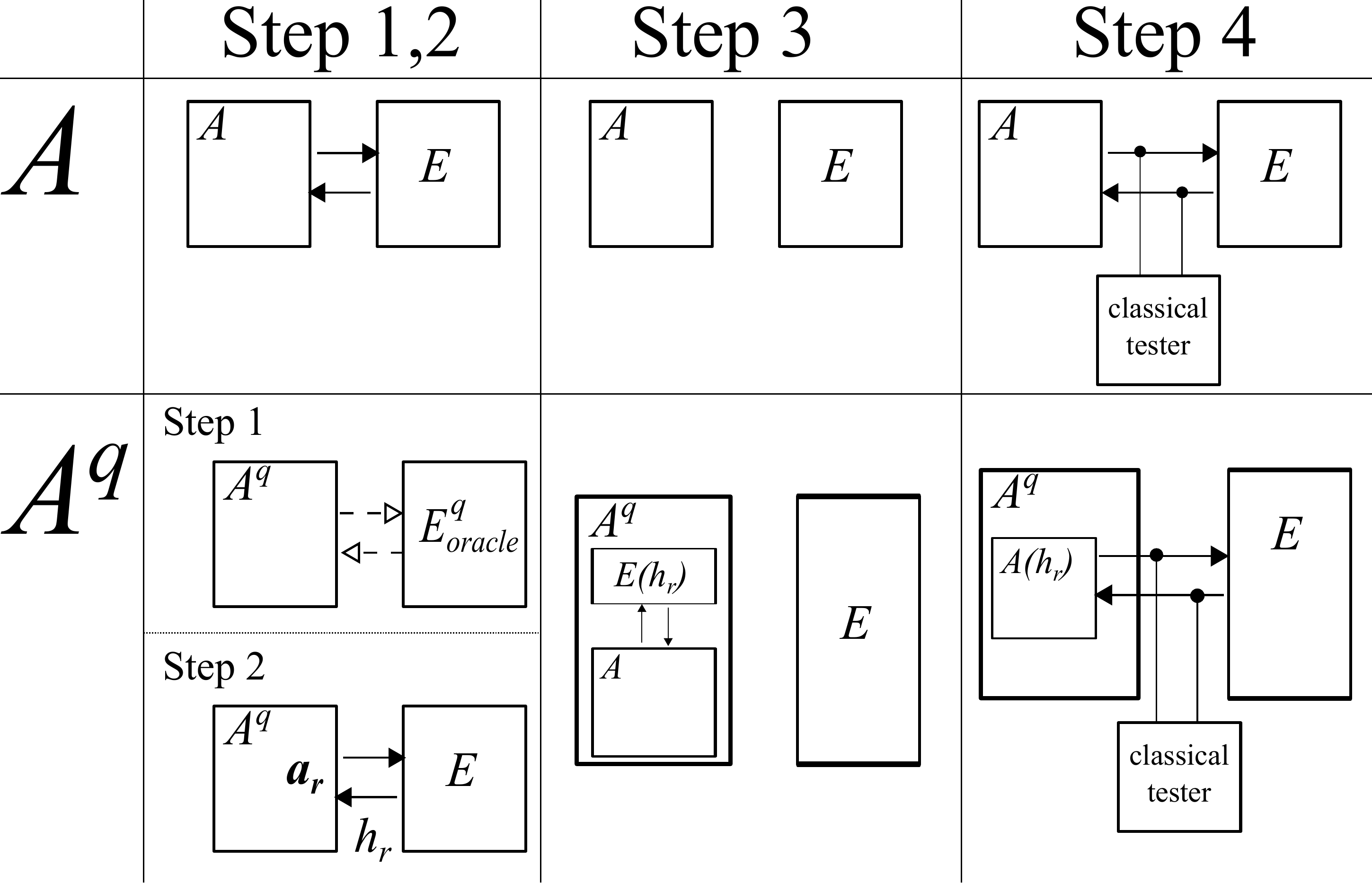}}

\caption{\label{construction}
The figure illustrates the differences between the agent-environment interaction for $A$ and the quantum-enhanced $A^q$. In Steps \ref{prep1} and \ref{prep2}, $A^q$ uses access to the oracular instantiation of $E,$ and obtains a winning sequence in, on average, a quadratically reduced number of interaction steps. At Step \ref{training}, $A^q$ simulates the agent $A$ internally, and `trains' this simulation to produce the sequence $h_{tot},$ derived from the winning sequence. During this time, there is no interaction between the agent and the environment, as this is `in between' time-steps. In Step 4, $A^q$ simulates $A(h_{tot})$ (using the obtained winning sequence $h_{win}$),  for the remainder of the interaction, now with the classical environment $E$. The interaction can be classically tested from this point on.
}
\end{figure}

We now briefly clarify the purpose of the steps in the construction. The construction is designed to guarantee improvement in luck-favoring settings. Steps \ref{prep1} and \ref{prep2} simply utilize Grover-like search to obtain (at least) one winning sequence of steps in the given environment, in time quadratically faster than would be possible for a classical agent. 
To understand the rest of the construction, we can ignore the quantum aspects and consider how one could utilize the knowledge of an agent $A$ given a winning sequence, without specifying the internal model. Step \ref{training} aims to achieve just that - it simulates an interaction with the agent $A,$ and resets the agent, until the desired sequence has been achieved. In quantum information terminology, the runs of the agent $A$ get post-selected to the winning branch. However, the number of interactions that have been experienced to this point are ($k \times\sqrt{n^M}$ times) larger than the length of the winning sequence ($M$). To compensate for this, and to put $A$ and $A^q$ on equal footing, this `postselection' is iterated on a larger scale - until the agent (by chance alone) reproduces the winning sequence $k \times\sqrt{n^M}$ times in a row. We have omitted this technical detail in the main text for clarity of presentation.
Alternatively to this, one can consider a broader definition of luck-favoring settings, where the two histories $h$ and $\tilde{h}$ (experienced by the `lucky', and `unlucky' agent respectively) may be of unequal lengths. But while we can argue that most reasonable environment-agent pairs are luck-favoring regarding the definition we have given, this will not be the case if the  sequences can arbitrarily differ in lenght.

This choice of the process of `training' a reinforcement learning model, given a winning sequence (or many winning sequences) is not crucial for our main point. However, regarding the optimization of the performance of the learning agent $A^q$, depending on how much is known about the learning model underlying $A,$ it should be chosen such that it maximizes the expected performance.  
To get further insight into the expected performance of $A$ versus $A^q$, consider the average configurations (relative to input-output behavior) of the agents $A$ and $A^q$ after the first $t$ steps.

Concerning agent $A^q$, after the time-step $t,$ and except with probability $O(\exp(-k))$, its behavior will be identical to the behavior of $A(h_{tot}),$ where $h$ is the history containing $(1+k\times\sqrt{n^M})$ successful move sequences glued together. 

%Due to the nature of Grover's search, the winning history $h$ is chosen uniformly at f from the set of sequences of winning histories of length $M_{max}$.

The configuration of the classical agent $A$, facing the same environment, is a bit more complicated, and what can be said is restricted by the fact that we do not specify the learning model of $A$.

Agent $A$ has also undergone $t$ interactions with the environment, that is, $(1+k\times\sqrt{n^M})$ complete epochs. 

The probability, however, of $A$ having seen at least one winning sequence (assuming there is no prior knowledge available to the agent) is upper bounded by the following expression:
\EQ{
P =1- \left( 1 - \dfrac{1}{n^M}   \right)\left( 1 - \dfrac{1}{n^M-1}   \right) \cdots \left( 1 - \dfrac{1}{n^M-k\times\sqrt{n^M}}   \right),
}
where we have taken into account the fact that the agent may (in the optimal case) never re-try a sequence which was not rewarded.
That expression further simplifies to 
\EQ{
P = \dfrac{k}{\sqrt{n^M}} + \dfrac{1}{n^M}
}
which decays exponentially to zero, for any fixed $k$, in $M$.
If we, for concreteness, set $k=M$ we have that both the probability $P$, and the failure probability of the quantum agent $O(\exp(-k)) = O(\exp(-M))$ decays exponentially  in $M$.
Thus, except exponentially small probability in $M$, the quantum agent will, from time-step $t$ onwards behave as 
$A(h_{tot}),$ where $h_{tot}$ has a maximal rate of rewards, whereas the classical agent will behave as   $A(h_{fail}),$ where $h_{fail}$ has not one rewarded percept.
Then, relative to any figure of merit $Rate$ which is increasing in the reward frequency (and depends only on the rewards) we have that 
$Rate(h_{tot}) > Rate(h_{fail})$.

Now, if the environment is luck-favoring, by Eq. (\ref{main-eq-luck}), from time-step $t$ onwards, the average performance of $A^q = A(h_{tot})$ will  beat the performance of $A(h_{tot})$ except with exponentially small probability, relative to the classical tester.

These observations form the first qualitative result, here given in full detail:

\TH\label{th1}
Let $E$ be a controllable environment, over action space $\mathcal{A}$,  thus it is, on the agent's demand, accessible in the form $E^q_{control}$. Moreover, let $E$ correspond to a deterministic, fixed-time $M$, single-win game, with a unique winning sequence of length $M$, for the period of $O(|\mathcal{A}|^M)$ time-steps (after which it no longer needs to be controllable, nor deterministic, fixed-time, single win).
Let  $A$ be a learning agent such that $(E, A)$ are luck-favoring for all histories, relative to some figure of merit $Rate(\cdot)$, which is increasing in the number of rewards in the history, and which only depends on the rewards.
Then there exists a quantum learning agent $A^q$ based on $A$ which outperforms $A$ in terms of $Rate(\cdot)$ and relative to a chosen sporadic classical tester.  
\HT

The above is the least one can establish. If we start specifying the scenario further, by e.g. fixing the $Rate(\cdot)$ to be an effective (normalized) counter of the rewards, then we can also consider the average number of interaction steps which the classical agent needs to perform (relative to the quantum agents $t =   k \times\sqrt{n^M}\times{M}+M$)  before the two agents can even in principle start achieving approximately equal behaviors in terms of the rate. Note that every sensible learning agent will, given a sufficient number of steps, start producing the winning sequence every subsequent game. In this case, the rate will be maximal for all such agents.
As we have clarified, the classical agent requires an average $t_c \in  O(n^M\times{M})$ interaction steps (so $O(n^M)$ complete epochs), before a rewarded sequence is seen even once, on average. Thus this establishes a reasonable lower bound on the order of the number of steps required for a classical agent to start approaching the performance of the quantum agent. This constitutes a quadratic improvement.
However, making such claims more formal requires further specifying the underlying learning model. Here, we wish to establish more general claims, and leave more specific analyses for future work.

Nonetheless, for concreteness, we can list examples of learning models, and task environments, where the quadratic improvement mentioned above is easy to argue. 

 If we additionally label each percept (think positions, or directions of optimal moves in maze problems), then many well-studied reinforcement learning models (e.g. Q-Learning~\cite{SuttonBarto98,2003_Russel}, Policy iteration~\cite{1990_Sutton} or the more recent Projective Simulation~\cite{2012_Briegel, 2014_Melinkov} model), together with the maze environment (with a unique winning path) do form luck-favoring pairs for all histories, so Theorem \ref{th1} applies. 
To further explain why this is the case (but without going into the details of these learning models), recall that in this single-win, bounded maximal time $(M)$ case, there is only one $M$-length history which has a reward. Moreover, a rewarding percept can only appear after exactly $M$ interaction steps, as the game is reset after each $M$ steps. Next, note that every history length, since it is an integer, can be written as $l\times M + q,$ with $q<M$, for some integers $l, q$.
In such a history the last $q$ percepts cannot be rewarding, so we can focus on histories of lengths $l \times M$. This can be interpreted as an $l$-fold concatenation of histories of lengths $M$. Each one of these $l$ sub-histories either has exactly one rewarding percept, or does not, and it does only if that sub-history is the unique winning sequence.
In the learning models we have mentioned, applied to such an environment, for every game where a winning sequence has been executed, the probability of executing the same winning sequence typically increases. This implies that for any two histories (independently of their length)  Eq. \ref{main-eq-luck2} holds. Moreover, if the environment does not change, it holds for all execution lengths , hence Theorem \ref{th1} does apply.

Going beyond the learning models we have mentioned, it is arguable that any learning model which is \emph{not} luck favoring with such a environment is a deficient learning model, as this would imply that the performance of the model (or the agent) does not monotonically improve as the agent encounters new short(er) paths.

In contrast, environments which are not luck-favoring with standard learning models are possible to concoct. 
Simplest examples include malicious environments that change the rules
 depending on the initial success of the agent. In this case, having a low efficiency in the exploration phase may be beneficial in the long run, but such scenarios are quite artificial.
In the next section, we will consider further generalizations of environments where a speed-up is possible, and in the process touch upon more reasonable (and more general) settings where being lucky may be not as advantageous, and comment how to deal with such settings.

\label{improvements}
\section{Generalizations}
\label{general}
Here we give examples of how oracles which correspond to multiply rewarding environments and stochastic environments can be utilized, and provide directions in which our approach can further be developed

Note that, in the main text, we have shown how stochasic environments can be mapped to oracles which present the probability of a reward being issued:
\EQ{
\ket{\ve{a}} \ket{0} \stackrel{U_{Stoch}}{\longrightarrow} \ket{\ve{a}} \ket{\tilde{\ve{\theta}_a}},
}
where $\tilde{\ve{\theta}_\ve{a}}$ is a representation of an approximation of the rewarding probability. Note, representing low probabilities is expensive, as each bit of the representation requires an additional layer of phase estimation. Importantly, $U_{Stoch}$ can be realized such that it is Hermitian, so self-inverse.
In section \ref{counting}, we have shown how counting oracles can be implemented, which realize the mapping

\EQ{
\ket{\ve{a}}_{\textup{I}} \ket{\ve{\bar{s}}}_{\textup{II}} \ket{y}_{\textup{III}} \stackrel{U_{Count}}{\longrightarrow}  
\ket{\ve{a}}_{\textup{I}} \ket{\ve{\bar{s}}}_{\textup{II}}  \ket{\Sigma \lambda \oplus y}_{\textup{III}},
}
where $U_{Count}$ is also self-inverse.

In both cases, we can thus use phase-kick back to amplitude-amplify \cite{2000_Brassard} sequences of actions satisfying a desired criterion, and note that relative to most figures of merit, the expected reward of a sequence in the stochastic environment has the same operational meaning as the total reward in the multiply rewarding deterministic environments.

Here, for instance, one can employ the methods of quantum optimization over discrete sets \cite{1996_Durr, 2005_Baritompa}.
In essence, in multiply rewarding environments (also stochastic environments) , we can choose a threshold of minimal (probability of) reward $p_{min}$, and perform the amplification of all amplitudes of action sequences which yield a reward (with probability) of at least $p_{min}$.
The number of environmental oracle calls will be $\sqrt{ N_{tot}/N_t} $ where $N_{tot} = |\mathcal{A}|^M$ is the total number of sequences and $N_t$ is the number of sequences satisfying the threshold criterion.
This can be achieved either by using a randomized Grover approach \cite{1998_Boyer} or using the optimal fixed-point approach of \cite{2014_Yoder}. 
In the latter case, we do not use phase kick back to mark the sequences but rather introduce an ancillary system and realize a `bit-flip' oracle, to put the abstract problem in the formulation given in \cite{2014_Yoder}.

As we have show in the main text, in the case of stochastic environments, we also must take the cost of finding the estimate of the probability of reward, and this incurs an additional cost of $O(1/p_{min})$ (note, in this setting $p_{min}$ is a probability). If the minimal probability is constant for a family of environments, then this is a constant cost as well.
In the multiply rewarding setting, we do not have this cost as no phase estimation is needed.
Thus in both cases (stochastic with a promise on the minimal relevant reward, and multiply rewarding case), we can obtain one sequence satisfying the criterion that the reward is above threshold (or occurs with probability above the threshold) quadratically faster than through classical interaction.

However, as in the basic case of single-reward deterministic environments, faster finding does not generically imply improved learning. Nonetheless, it is easy to identify some settings where this follows immediately in luck-favoring settings. One example are settings where the large rewards are scarce, and nearly all moves of the agent yield a (bounded) small reward. More formally, whenever high (or high-probability rewards) occur only for a constant number (or log-sized number, in the total number of sequences) of sequences, whereas the low rewarding sequences occur for a fraction of sequences, we obtain an improvement in learning. This is achieved by using analogous constructions as for the single-win deterministic case.
\subsection{Stochastic environments with structural dependence}
The stochastic environments where our proposed constructions help are those where  action sequences can a-priori have a higher or lower probability of being rewarded.
However, there are settings in which this is manifestly not the case.
Consider the settings where each percept is chosen uniformly at random by the environment, but once this choice is made, there is only one correct action for the agent, different for each choice of the environment.
In this picture, the agent is to correctly respond to a random percept of the environment.
This problem becomes more interesting when the reward is issued only after $M$ steps (this is a modification of the contextual bandit problem, or the invasion game in \cite{2012_Briegel}). In this case, a search over just the action space does not reveal useful information, as any sequence of actions (if we trace over the percepts of the environment) is equally likely to yield a reward.

We can provide useful oracles for this case as well.

We start with the purified picture of the maps of the environment as given in the construction for the basic stochastic-variant oracle, albeit rewritten to a form useful for this case:

\EQ{
\ket{\ve{a}}_{\textup{I}} \otimes \ket{0}_{\textup{II}}\otimes \ket{0}_{\textup{III}} \stackrel{S_E}{\longrightarrow} 
\ket{\ve{a}}_{\textup{I}} \otimes  \sum_{\ve{s}} \sqrt{p(\ve{a})} \ket{\ve{s},\ve{s}}_{\textup{II}}\otimes \ket{\lambda(s,a)}_{\textup{III}} ,
}
where the second part of the state $\ket{\ve{s},\ve{s}}$ purifies the otherwise stochastic dynamics, and we, for simplicity, assume the percept choice of the environment is independent from the actions. This too can be relaxed, in principle.

Re-writing this, we obtain

\EQ{
\ket{\ve{a}}_{\textup{I}} \otimes \ket{0}_{\textup{II}}\otimes \ket{0}_{\textup{III}} \stackrel{S_E}{\longrightarrow} 
\ket{\ve{a}}_{\textup{I}} \otimes  \left( \sum_{\ve{s}, \lambda(s,a)=0} \sqrt{p(\ve{a})} \ket{\ve{s},\ve{s}}_{\textup{II}}\otimes \ket{0}_{\textup{III}}  + \sum_{\ve{s}, \lambda(s,a)=1} \sqrt{p(\ve{a})} \ket{\ve{s},\ve{s}}_{\textup{II}}\otimes \ket{1}_{\textup{III}} \right).
}
Note that applying a Pauli-Z to the register $\textup{III}$ is equivalent to reflecting about the state
\EQ{
\ket{\pi_{target}} =  \gamma\sum_{\ve{a}} 1/(\sqrt{|\mathcal{A}|})\ket{\ve{a}} \otimes \sum_{\ve{s}, \lambda(s,a)=1} \sqrt{p(\ve{a})} \ket{\ve{s},\ve{s}}_{\textup{II}}\otimes \ket{1}_{\textup{III}},
}
where $\gamma$ is a normalization factor.
Note, in the state above, the second register contains only the encodings of all percept sequences $\ve{s}$ which, in conjunction with the action sequence $\ve{a}$ yield a reward.
Finally, assume also that the overall mapping is self-inverse (and as we have clarified earlier, there always exist realizations of the same environment where this can be enforced).
Now note that
$S_E U_{\ve{a}}(Z_{\textup{I}} Z_{\textup{II}} Z_{\textup{III}})U_{\ve{a}}^\dagger S_E$, where $U_{\ve{a}}$ is such that
$U_\ve{a} \ket{0} =\sum_{\ve{a}} 1/(\sqrt{|\mathcal{A}|})\ket{\ve{a}}, $
 constitutes a reflection about the state
\EQ{
\ket{\pi} =S_E \sum_{\ve{a}} 1/(\sqrt{|\mathcal{A}|})\ket{\ve{a}}\ket{0}\ket{0}.
}

Also note that, on the subspace spanned by $ \{ \ket{\pi_{target}},   \ket{\pi_{target}^\bot}\}, $ where
\EQ{
 \ket{\pi_{target}^\bot} = \sqrt{1-|\gamma|^2} \sum_{\ve{a}} 1/(\sqrt{|\mathcal{A}|})\ket{\ve{a}} \otimes \sum_{\ve{s}, \lambda(s,a)=0} \sqrt{p(\ve{a})} \ket{\ve{s},\ve{s}}_{\textup{II}}\otimes \ket{0}_{\textup{III}},
}
the operator $Z_{\textup{III}}$ (Pauli-Z applied to the third register) constitutes a reflection about the state $\ket{\pi_{target}}.$ Since both operations are implementable by the agent, it can perform amplitude amplification, amplifying the amplitudes of the state $ \ket{\pi_{target}}$, (approximately) reaching it using $O(|\bra{\pi} \pi_{target}\rangle|^{-1})$ oracular calls to the environment.
By measuring $\ket{\pi_{target}},$ a quantum agent can learn one pair $\ve{a},\ve{s}$ which yield a reward.

The classical cost of the same process would be quadratically worse.

This can then be iterated $O(|\bra{\pi} \pi_{target}\rangle|^{-1})$ many times before any classical agent finds even one pair.
The quantum agent, at that point, has $O(|\bra{\pi} \pi_{target}\rangle|^{-1})$ many samples from a conditional distribution of the actions and percept pairs which get rewarded.
This sample set can then be used to train the classical agent, similar to the approach we used in the proof of the main theorem. Effectively, the sample suffices for a partial representation of the actual environment. We can use this representation for the repeat-until-success approach again, post-selecting the interaction between the simulation of the environment and the agent, and allowing only runs in which the agent responds in a manner compatible with the sample set. Note, only in this case can we faithfully simulate the environment.

What this realizes is again a (particularly) lucky agent in the sense of definition of luck-favoring environments.
Then, by the same arguments as before, such a trained agent will outperform a classical agent in all luck favoring settings. We leave the details of this construction for future work.

\end{widetext}

\end{document}